\documentclass[11pt, a4paper, openany]{book} % General settings 
\usepackage[german,english]{babel} 
\usepackage{rotating,booktabs,multirow}
\usepackage{comment} 
\usepackage[utf8]{inputenc}
\usepackage[T1]{fontenc}
\usepackage{amsmath,amsfonts,amsthm}
\usepackage[toc,page]{appendix}
\usepackage{algorithm}
\usepackage[noend]{algpseudocode}
\usepackage{natbib}

\usepackage{titlesec}

\titleformat{\chapter}[display]
    {\normalfont\huge\bfseries}{\chaptertitlename\ \thechapter}{20pt}{\Huge}
\titlespacing*{\chapter}{0pt}{0pt}{20pt}

\usepackage{geometry} % Necessary package for defining margins

\usepackage{setspace} % Required for spacing
\usepackage[T1]{fontenc} % Output font encoding for international characters
\usepackage[utf8]{inputenc} % Required for inputting international characters

\usepackage{XCharter} % Use the XCharter font

\usepackage{fancyhdr} % Package is needed to define header and footer
\usepackage{filecontents}
\usepackage[]{graphicx}

\begin{document}

\begin{titlepage}
\begin{center}
Rheinische Friedrich-Wilhelms-Universität Bonn
		 
Master Programme in Economics
		
		\vspace{1in}
		 {\large \bfseries \onehalfspacing Term Paper for Research Module in\par Econometrics \& Statistics}
		\vspace{1in}
		
		{\LARGE \bfseries \onehalfspacing Weak Instrumental Variables: \par Limitations of Traditional 2SLS\par and \par Exploring Alternative Instrumental Variable Estimators}
		\vspace{1in}
		
		{\large Submitted by:}
		
		{\Large Aiwei Huang\par Madhurima Chandra \par Laura Malkhasyan\par}
		
		\vspace{1in}
		
			{\large \linespread{2} Supervised by: \par \Large JProf. Dr. Dominik Liebl}

		\begin{center}
			{\large February 05, 2020}
		\end{center}
		\end{center}
	\end{titlepage}

%---------------------------------------------------------------------------------------

\tableofcontents

%---------------------------------------------------------------------------------------

%----------------------------------------------------------------------------------------
% Introduction
%----------------------------------------------------------------------------------------

\setcounter{page}{1} % Sets counter of page to 1
\chapter{Introduction} Instrumental variables estimation has gained considerable traction in recent decades as a tool for causal inference, particularly amongst empirical researchers. However, this has also highlighted the importance of taking a deeper look at the theoretical properties underlying one's estimator of choice, whether it is instrumental variable estimator or any other estimator. A case-in-point is the well-known 1991 quarter-of-birth study on returns to schooling estimates by \cite{angrist1991does} and the equally famous rebuttal of their results published by \cite{bound1995problems}.

\par The \cite{bound1995problems} critique of the quarter-of-birth study in essence formed the starting point of the weak instruments literature. They provided simulation evidence that the instruments used by \cite{angrist1991does} were very weak and this in turn led to misleading results. Interestingly, their study was not the first time that the poor finite-sample behavior of the instrumental variables estimator (two-stage least squares estimator in particular) was discussed. Previously, numerous authors had presented results on the magnitude of the finite-sample bias of the 2SLS estimator towards OLS, such as \cite{nagar1959bias}, \cite{basmann1960asymptotic}, \cite{richardson1968exact} and \cite{sawa1969exact}.
However, their critique was the first to suggest some practices for the detection of possible presence of weak instruments (reporting first-stage F-statistic and the $R^2$ of the first stage regression), which have since become standard practice for using instrumental variables in empirical studies.

\par The literature on weak instruments have since developed considerably and continues to do so. \cite{staiger1997stock} showed that conventional asymptotic approaches fail to provide good approximations for weak instruments, and formalized the problem of weak instruments by introducing `weak-instrument asymptotics' which mimics the situation of weak instruments better. \cite{stock2002survey} provide a comprehensive survey of the weak instruments literature, wherein they emphasize that the problem of bias of the two-stage least-squares estimator is not
solely a small sample problem. It affects estimates carried out on large sample sizes as well, as is the case with the quarter-of-birth study which has a dataset of around 300,000 observations.

\par Further, given the poor finite-sample properties of the two-stage least squares estimator, numerous alternate estimators have also been proposed. These include the USSIV estimator suggested by \cite{angrist1995split}, Fuller-k estimator by \cite{fuller1977some}, bias-adjusted 2SLS estimator by \cite{donald2001choosing}, JIVE estimators by \cite{angrist1999jackknife} and \cite{blomquist1999small}. Another estimator, the LIML was formalized by \cite{anderson1949estimation}. In our paper, we take a deeper look at two of these estimators - JIVE and LIML estimators.

\par We introduce the two conditions for instrumental variables estimation, the first of which will be a recurring theme in our paper, since it connects to the problem of weak instruments. Consider the standard population regression model:
\begin{align*}
Y_i &= \beta_0 + \beta_1X_i + \varepsilon_i, \: \: \: \: i = 1,2,...n 
\end{align*}
where $\varepsilon_i$ is the error term representing omitted factors that determine $Y_i$. Variables correlated with the error term are termed endogenous variables and those uncorrelated with the error term are exogenous. A valid instrument $Z_i$ must satisfy two conditions:
\begin{itemize}
    \item Instrument Relevance: corr ($Z_i$, $X_i$) $\neq 0.$\\
    It is the implications of this condition, specifically the strength (or weakness) of this correlation between the instrument and the endogenous regressor, that we are interested in. Instruments that explain little of the variation in X are called weak instruments. A precise definition is introduced in section 3.1.
    \item Instrument Exogeneity: corr ($Z_i$, $\varepsilon_i$) = 0\\
    Note that this condition cannot be statistically tested, since it involves the covariance between $Z_i$ and the unobserved $\varepsilon_i$.
\end{itemize}
\par If number of instruments (K) equals the number
of endogenous regressors (L) we have the just or exactly identified case.  If the number of instruments exceeds the number of endogenous regressors, K $\geq$ L, then we have the over-identified case. 

\par This paper makes three contributions. First, we provide a detailed theoretical discussion on the properties of the standard two-stage least squares estimator in the presence of weak instruments, and introduce and derive two alternative estimators. Second, we conduct Monte-Carlo simulations to compare the finite-sample behavior of the different estimators, particularly in the weak-instruments case. Third, we apply the estimators to a real-world context; we employ the different estimators to calculate returns to schooling.

\par The rest of the paper is sectioned as follows. Chapter 2 presents a derivation of the 2SLS estimator and its limitations in the weak instruments case. Chapter 3 discusses methods to test for weak instruments. Chapter 4 presents the alternative estimators (JIVE and LIML). Chapter 5 presents the simulation study. Chapter 6 presents the application to returns to schooling. Chapter 7 concludes.

\chapter{Two-Stage Least Squares and its Limitations}

\section{Basic Model}
Throughout the paper, unless stated otherwise, we consider the following basic model with two equations, following the notation of \cite{angrist1999jackknife}:
\begin{align*}
Y_i &= X_i\beta + \varepsilon_i \\
X_i &= Z_i\pi + \eta_i
\end{align*}
$Y_i$ is scalar. $X_i$ is $1\times L$ row vector of potentially endogenous regressors. 
$Z_i$ is $1\times K$ row vector, with $K\geq L$. 
$K-L$ is the number of over-identifying restrictions. In matrix notation:\\
\begin{align}
\underset{(N\times 1)}{\mathbf{Y}} = \underset{(N\times L)}{\mathbf{X}}\beta + \underset{(N\times 1)}{\varepsilon} \\
\underset{(N\times L)}{\mathbf{X}} =  \underset{(N\times K)}{\mathbf{Z}}\pi + \underset{(N\times L)}{\eta}
\end{align}
where (2.1) denotes the structural equation, (2.2) is the first stage. $\mathbf{Y}$ and $\varepsilon$ are $N \times 1$ vectors. 
$\mathbf{X}$ and $\eta$ are $N \times L$ matrices, and $\mathbf{Z}$ is $N \times K$ matrix. If the vectors of regressors and that of instruments have $M$ common elements, then $M$ columns of the $N \times L$ matrix $\eta$ have their elements as zero.
\par The following assumptions hold for our model:
\begin{enumerate}
\item Conditional on instruments $Z_i$, the error term $\varepsilon_i$ has expectation zero ($E[\varepsilon_i|Z_i]=0$) and variance $\sigma^2$.
\item $E[\eta|\mathbf Z]=0$ and $E[\eta_i^\prime\eta_i|\mathbf Z]=\Sigma_\eta$, with rank $L-M$.
\item $E[\varepsilon_i\eta_i^\prime|\mathbf Z]=\sigma_{\varepsilon\eta}$, where $\sigma_{\varepsilon\eta}$ is an L-dimensional column vector.
\item All observations of ($Y_i, X_i, Z_i$) are independent and identically distributed.
\end{enumerate}

\section{Concentration Parameter}
The Concentration Parameter $\mu^2$, given by:
        \begin{equation}
  \mu^2 = \frac{\pi' Z' Z \pi}{\sigma_{\eta}^2}
        \end{equation}
measures the strength of the instruments. It is unitless. An important question to be addressed at this point is that since $\pi$, and thus $\mu^2$, is unknown, how is the researcher supposed to know whether $\mu^2$ value is low enough for the instrument under study to be weak? This is addressed by the fact that the concentration parameter can be interpreted in terms of the first stage F statistic, discussed in Section 3. If the sample size is large, $E(F) \cong \mu^2/K + 1$, and thus $F-1$ can be treated as an estimator of $\mu^2 /K$, which gives us a convenient way to test for weak instruments.

\par The concept of $\mu^2$ is essentially the starting point for understanding the weak instruments literature. \cite{rothenberg1984approximating} showed that the concentration parameter plays the role we commonly associate with the sample size or number of observations: as $\mu^2$ becomes large, the normal distribution becomes a good approximation for the 2SLS distribution. Likewise a small concentration parameter leads to a non-normal distribution of 2SLS, and the estimator becomes biased. Thus $\mu^2$ can be interpreted as the effective sample size.

\section{Two-Stage Least Squares Estimator}

It is helpful to characterize the 2SLS and (later on the jackknife estimator as well), as a feasible version of the ideal but infeasible instrumental variables estimator. Adhering to the notation defined in section 2.1, we derive the formula for the 2SLS estimator using this characterization. The derivation is adopted, with modifications to the notation, from \cite{brucehansen2019} and \cite{angrist1999jackknife}. 
\par First, we estimate the ideal estimator $\hat\beta_{OPT}$ using the optimal instrument $Z\pi$, by ordinary least-squares:
\begin{equation}
\hat\beta_{OPT}= ((\mathbf{Z}\pi)'\mathbf{X})^{-1}((\mathbf{Z}\pi)'\mathbf{Y})
\end{equation}
$\pi$ is unknown however, and so the optimal estimator $\hat\beta_{OPT}$ is infeasible.

We use an estimate of $\pi$ instead, and denote it  by $\hat\pi$. $\hat\pi$ is estimated from the reduced form regression which yields $\hat\pi = (\mathbf{Z}'\mathbf{Z})^{-1}(\mathbf{Z}'\mathbf{X})$. Substituting $\hat\pi$ into (2.4),
\begin{equation}
\hat\beta_{2SLS}= ((\mathbf{Z}\hat\pi)'(\mathbf{Z}\hat\pi))^{-1}((\mathbf{Z}\hat\pi)'\mathbf{Y})
\end{equation}
\begin{equation}
\hat\beta_{2SLS}= ((\mathbf{Z}(\mathbf{Z}'\mathbf{Z})^{-1}(\mathbf{Z}'\mathbf{X}))'(\mathbf{Z}(\mathbf{Z}'\mathbf{Z})^{-1}(\mathbf{Z}'\mathbf{X}))^{-1}((\mathbf{Z}(\mathbf{Z}'\mathbf{Z})^{-1}(\mathbf{Z}'\mathbf{X}))'\mathbf{Y})
\end{equation}
\begin{equation}
\hat\beta_{2SLS}= ((\mathbf{X}'\mathbf{Z})(\mathbf Z'\mathbf Z)^{-1}\mathbf Z'\mathbf X)^{-1}\mathbf X'\mathbf Z(\mathbf Z'\mathbf Z)^{-1}\mathbf Z'\mathbf Y
\end{equation}
This is the two-stage least squares estimator.
$\hat\beta_{2SLS}$ can be thought of as an estimator with the constructed instrument $(\mathbf{Z}\hat\pi)$ where $\hat\pi = (\mathbf{Z}'\mathbf{Z})^{-1}(\mathbf{Z}'\mathbf{X})$.
\par
Representing (2.7) in terms of the projection matrix  $P_{Z}=\mathbf Z(\mathbf Z'\mathbf Z)^{-1}\mathbf Z'$:
\begin{equation}
\hat{\beta }_{2SLS}=((\mathbf X'P_{Z}\mathbf X)^{-1}\mathbf X'P_{Z}\mathbf Y)
\end{equation}
The two-stage least squares is by far the most widely used estimator for instrumental variables estimation, however, as we briefly discussed in chapter 1, it has certain concerning limitations, which we expand upon in the next section.

\section{Limitations of 2SLS: Bias}
In the weak instruments case, the 2SLS estimator fails to provide unbiased, reliable estimates. As shown in Section 2.3, we arrived at the 2SLS estimator using an estimate of $\pi$ since the true value of the first-stage coefficients is unknown, which implies a certain amount of
over-fitting of the first-stage equation, leading to bias in the
direction of the expectation of the OLS estimator of $\beta$.

\par There is considerable  theoretical research noting the poor finite-sample properties of instrumental variables estimator. Results on the magnitude of this bias have been provided by \cite{nagar1959bias}, \cite{richardson1968exact}, \cite{sawa1969exact} and \cite{buse1992bias}. Further, as shown by noteworthy results of \cite{bound1995problems}, with weak instruments particularly, the finite sample behavior worsens and the estimator can become biased as well as inconsistent.

\subsection{Bias in Single Instrument case}
To visualize the problem intuitively, first let us consider the simple case of a single (weak) instrument and endogenous regressor:

        $$Y_i = X_i\beta + \varepsilon_i$$
$$X_i = Z_i\pi_1 + \eta_i$$

In this case, the 2SLS estimator is simply given by the ratio of the two covariances: 
\begin{equation}
    \hat\beta_{2SLS}=\frac{\sigma_{Y_iZ_i}}{\sigma_{X_iZ_i}}
\end{equation}
However, given the fact that our instrument is very weak,
$\sigma_{X_iZ_i}=0.$ 
Hence, $\hat\beta_{2SLS}$ does not exist.
\par In the simplest case, the 2SLS estimator is just the ratio of two covariances, and with weak instruments, the 2SLS or general instrumental variables estimator does not exist.

\subsection{Approximate Expression of Bias in Multiple-Instruments case}
In this section we provide two different expressions approximating the bias of the two-stage least squares estimator in the general case and show how it centers around ordinary least squares. 
\par Defining the `relative bias' of 2SLS to be its bias relative to the
inconsistency of OLS, \cite{buse1992bias} derived an expression for the approximate bias of $\hat\beta_{2SLS}$ using power series approximations. This result holds even when the errors are not normally distributed: 
\begin{equation}
  \frac{\sigma_{\varepsilon\eta }(K-2)}{\pi'\mathbf Z'\mathbf Z\pi}  
\end{equation}

where N is the sample size and K is the number of excluded instruments.
This expression is approximately inversely proportional to $\mu^2/(K-2)$, as shown:

\begin{equation}
\frac{\sigma_{\varepsilon\eta}{\sigma_\eta^2}(K-2)}{{\sigma_\eta^2}\pi'\mathbf Z'\mathbf Z\pi} = \frac{\sigma_{\varepsilon\eta}(K-2)}{{\sigma_\eta^2}{\mu^2}}  
\end{equation}
where $\mu^2$ denotes the concentration parameter (measure of the strength of the instrument as discussed in Section 1.). Note that in the above equation, the term $\frac{\sigma_{\varepsilon\eta }}{\sigma_{\eta }^{2}}$ approximately equals the asymptotic bias of the OLS estimator when the instrument explains little of the variation of $\mathbf X$. Hence, from equation (2.10) it is clearly seen that for $K>2$, the bias of the 2SLS estimator relative to OLS is inversely proportional to the concentration parameter, hence, weaker the instrument(s), lower is the concentration parameter, and higher is the bias of 2SLS towards the OLS estimate.

\par It can be seen from the above formulation that increasing the number of instruments with the explanatory power remaining constant, causes the relative bias of the 2SLS to only increase.

\par Before moving on to the second expression, it is important to briefly introduce alternate asymptotic representations that are generally used for the weak instruments case.
As discussed by \cite{stock2002survey}, for weak instruments, conventional asymptotic approximations to finite-sample distributions are quite poor. Two alternate asymptotic methods commonly employed are \textbf {weak-instrument asymptotics} (involving a sequence of models chosen to keep $\mu^2/K$ constant as sample size $N \rightarrow \infty$) pioneered by \cite{staiger1997stock} and \textbf {many-instrument or group asymptotics} (involving sequence of models with fixed instruments and normal errors, where $K$ is proportional to $N$ and $\mu^2/K$ converges to a constant finite limit), first proposed by \cite{bekker1994alternative}. In the 1995 working paper version of \cite{angrist1999jackknife}, it was referred to as group asymptotics.

\par We present an expression for the approximate bias of the 2SLS estimator using group asymptotics, wherein we let
the number of instruments grow proportional to rate of the sample size.
This keeps the instruments weak. The detailed derivation of this expression is provided in the appendix A.1.
\begin{equation}
    E[\hat{\beta }_{2SLS}-\beta]\approx \frac{\sigma_{\varepsilon\eta }}{\sigma_{\eta }^{2}}\frac{1}{F+1}
\end{equation}
where F is the population analog of the F-statistic for the joint significance of the instruments in the first-stage regression. $\frac{\sigma_{\varepsilon\eta }}{\sigma_{\eta }^{2}}$ approximately equals the asymptotic bias of the OLS estimator. Given a weak first-stage (weak instruments case), $F\rightarrow0$, and we can see that the bias approaches $ \frac{\sigma_{\varepsilon\eta }}{\sigma_{\eta }^{2}}$. With a strong first-stage, $F\rightarrow \infty$ and then the 2SLS bias goes to 0.

\subsection{Inconsistency of the 2SLS estimator}
In this section we relate the two conditions for instrumental variables estimation to the consistency property of the 2SLS estimator. If the weak correlation (between the instrument(s) and the endogenous variable) is coupled together with even a small violation of the second condition of instrumental variables estimation (which is the instrument exogeneity condition) then we have an inconsistent estimator. This insight was first discussed by \cite{bound1995problems}.

To represent the problem discussed above, let us consider the probability limit of the 2SLS estimator:

\begin{equation}
plim \hat{\beta} _{2SLS}=\beta + \frac{\sigma _{\mathbf {\hat X},\varepsilon }}{\sigma _{\mathbf {\hat X}}^{2}}
\end{equation}

where $\mathbf {\hat X}$ is the projection of $\mathbf X$ onto $\mathbf Z$, and ${\sigma _{\mathbf {\hat X},\varepsilon }}$ is the covariance between $\mathbf {\hat X}$ and $\varepsilon$.

From equation (2.13) we can intuitively understand that if $\sigma _{\mathbf {\hat X}}^{2}$ is small, which implies that we have a weak instrument, then as long as ${\sigma _{\mathbf {\hat X},\varepsilon }}$ is zero the estimator will be consistent. However, suppose we have a weak instrument  and also, ${\sigma _{\mathbf {\hat X},\varepsilon }}$ is small but non-zero, then even that small correlation between $\mathbf Z$ (and thus, $\mathbf {\hat X}$) and the structural error term $\varepsilon$ can lead to a large inconsistency.

Hence, with weak instruments, even moderate correlation between instrument and structural error term can magnify the inconsistency of IV estimator. 

\chapter{Testing for Weak Instruments}
Testing for presence of weak instruments is, at the time of writing, an active field of research. For a detailed overview, see \cite{stock2002survey}. For the purpose of our study, we limit our attention to two tests - the widely-used first-stage F-statistic and the Anderson-Rubin Test, which has gained resurgence in recent years in light of new developments in instrumental variables research. 

\section{Defining the `Weakness' precisely}
\cite{stock2002testing} posit that the definition of weak instruments depends on the inferential task to be carried out, and cannot be resolved in the abstract. One approach is to define a set of instruments to be weak if $\mu^2/K$
is small enough that inferences based on conventional normal
approximating distributions are misleading. For instance, if a researcher wants their 2SLS estimate bias to be small, one measure
of whether an instrument(s) is strong is whether $\mu^2/K$ is
large enough such that the 2SLS relative bias (relative to the bias of ordinary least squares) is below a certain threshold, for example the relative bias is below 10\%. Hence, to be deemed a `weak' instrument, the 2SLS estimate using that instrument should have relative bias above 10\%. The definition we discussed (and use for our simulation) is based on relative bias, another definition (for instance on size of test) may result in a different cut-off value.

\section{First Stage F-statistic}
The first-stage F-statistic is the F-statistic testing the hypothesis that the coefficients on the instruments equal zero ($\pi=0$) in the first stage of two stage least squares. 
\cite{stock2002testing} show that the definition of weak instruments discussed above implies a threshold value for $\mu^2/K$, under weak asymptotics. A weak instrument will have a $\mu^2/K$ value (and hence, an F-statistic value, since F$-$1 can be treated as an estimator of $\mu^2/K$ as discussed in Section 2.2) lower than the threshold.
For the case of a single endogenous regressor, \cite{staiger1997stock} provide a rule-of thumb threshold of 10: a value less than 10 indicates that the instruments are weak, in which case the 2SLS estimator is biased and 2SLS t-statistics and confidence intervals are unreliable.

\cite{stock2002survey} provide a table listing critical values of the first-stage F-statistic such that the relative bias of 2SLS estimates is greater than 10\%, for different numbers of instruments. The authors arrived at those critical values based on weak-instrument asymptotic approximations. We include a subset of this table (which is relevant for our simulations) as Table B.1 in appendix B for reference.

\section{Anderson-Rubin Test}

The AR test is a hypothesis test that has the property of being valid whether instruments are strong, weak or even irrelevant ($\pi=0$). It tests the null hypothesis $\beta$ = $\beta_0$ using the statistic. It was proposed by \cite{anderson1949estimation}.

\begin{equation}
  AR(\beta) = \frac{(\mathbf Y- \mathbf X\beta)' P_z (\mathbf Y-\mathbf X\beta)/K}{(\mathbf Y-\mathbf X\beta)' \mathbf M_Z (\mathbf Y-\mathbf X\beta)/(N-K)}
        \end{equation}
        
One definition of the LIML estimator is that it minimizes
$AR(\beta)$.
With fixed instruments and normal errors, the quadratic
forms in the numerator and denominator of (3.1) are independent
chi-squared random variables under the null hypothesis,
and $AR(\beta_0)$ has an exact $F_{K,T - K}$ null distribution. Under the more general conditions of weak-instrument asymptotics, $AR(\beta_0)$ $\xrightarrow{\text{d}}$ ${\chi_k}^2/K$ under the null hypothesis, regardless of the
value of $\mu^2/K$. Thus the AR statistic provides a fully robust
test of the hypothesis $\beta$ = $\beta_0$.
\par The set of
values of $\beta$ that are not rejected by a 5\% Anderson–Rubin test will constitute a 95\% confidence
set for $\beta$. The logic behind the Anderson–Rubin statistic is that it
never assumes instrument relevance, and the AR confidence set will have a
coverage probability of 95\% in large samples, regardless of the strength or weakness of instruments.
In light of the importance given to the problem of weak instruments in recent years, this test has gained traction among econometricians, who increasingly advocate for its use for robust inference with weak instruments (See \cite{staiger1997stock}). Particularly, recent research has shown the AR confidence set to be optimal in the single-endogenous-regressor just-identified setting.
\chapter{Alternative Estimators}
To tackle the problem of finite-sample bias of IV, which as we have shown, is particularly problematic in the presence of weak instruments. In this section we present a detailed overview of two alternative estimators which in theory exhibit better finite sample properties. 
\par Both of these estimators, the Jackknife IV estimator and the LIML estimator, fall under the broader class of k-class estimators. These estimators partially robust ie less sensitive to weak instruments; they are more reliable in comparison to 2SLS estimates.
\section{Jackknife Instrumental Variables Estimator}
\textbf For an intuitive sense of the functioning of the jackknife instrumental variables estimator, consider the term `jackknife' as presented in statistical literature. The  jackknife estimator of a parameter is found by leaving out each observation from a dataset, calculating the estimate and computing the average of these calculations. Given a sample of size $N$, the jackknife estimate is found by aggregating the estimates of each sub-sample of size $(N-1)$. In a similar vein, the estimator we describe in this section replaces the usual fitted values from the reduced form regression of the two-stage least squares by `omit-one' fitted values.
\par The key feature of the JIVE estimator is that it eliminates the correlation between the fitted values and the structural equation errors, as explain in detail below. The fitted value of the standard 2SLS estimator is only asymptotically independent of the structural error ($\varepsilon_i$), but the JIVE is independent even in finite samples.
\par The derivation we present in this section is adopted from \cite{angrist1999jackknife}. To derive the expression formally, consider again the expressions for $\hat\beta_{OPT}$ and $\hat\beta_{2SLS}$ from section 2.2, equations (2.4) and (2.5). $\hat\beta_{OPT}$ employed the optimal instrument $\mathbf Z\pi$ and $\hat\beta_{2SLS}$ used an estimate of $\mathbf Z\pi$, denoted by $\mathbf Z\hat \pi$. Now we introduce the new estimator for $\beta$ by using a different estimate of the optimal instrument, $Z_{i}\tilde{\pi}$. 

\par From the formulation of approximate bias of the 2SLS estimator we presented in section 2.4.2 (equation 2.11), it is clear that increasing the number of instruments while keeping the explanatory/predictive power of the instruments constant, leads to an increase in the bias of $\hat\beta_{2SLS}$. However, we can see that for the optimal estimator, increasing the number of instruments while keeping $Z_i\pi$ fixed will have no effect on the properties of $\hat\beta_{OPT}$. Thus, for finite samples in the presence of many instruments, $\hat\beta_{2SLS}$ fares much worse than $\hat\beta_{OPT}$. 
\par Recall the representation of the 2SLS estimator using the projection matrix $P_Z$ from Section 2 (equation 2.8). We can see that the first-stage fitted values $\mathbf Z\hat{\pi}$ can be written as:

\begin{equation}
\mathbf Z\hat{\pi} = P_Z\mathbf X = \mathbf Z \pi + P_Z\eta
\end{equation}

The term $P_Z\eta$ in the above equation is correlated with the error term of the first-stage $\eta$ (see equation 1.2) and hence with the structural error term $\varepsilon$. Put differently, since $\hat \pi$ is estimated on the full sample which includes the $i$th observation, it is correlated with $\eta_i$, which is correlated with $\varepsilon_i$. This correlation is given by:

\begin{equation}
E[\varepsilon_{i}Z_{i}\hat{\pi}]
=E[E[\varepsilon_i Z_i\hat\pi|\mathbf Z]]
=E[Z_i(\mathbf Z^\prime \mathbf Z)^{-1} Z_i^\prime] \cdot E[\varepsilon_i \eta_i|\mathbf Z]]
=E[Z_i(\mathbf Z^\prime \mathbf Z)^{-1} Z_i^\prime \cdot \sigma_{\varepsilon\eta}^\prime]
=(K/N)\cdot\sigma_{\varepsilon\eta}'
\end{equation}

Due to this correlation between $\hat \pi$ and $\varepsilon_i$, $\hat\beta_{2SLS}$ is biased for $\beta$. While the correlation disappears asymptotically, it holds implications in finite samples. 

\par In devising the new instrument, we attempt to keep the correlation in (4.2) equal to zero. The problem stems from $\hat \pi$ being estimated on the full sample which includes the $i$th observation. For the new estimator which has the constructed instrument $Z_{i}\tilde{\pi}$, $\tilde\pi$ is estimated not on the full sample but on the sample with the $i$th observation removed. Therefore, the new estimated instrument will be independent of $\varepsilon_i$ even in finite samples.

The $i$th row of the estimated instrument for 2SLS, $\mathbf Z\hat{\pi}$, where $\hat{\pi}=(\mathbf Z'\mathbf Z)^{-1}(\mathbf Z'\mathbf X)$, is given by
\begin{equation}
Z_{i}\hat{\pi}= Z_{i}({\mathbf Z}'\mathbf Z)^{-1}(\mathbf Z'\mathbf X)
\end{equation}
 
Remove the $i$th row from the matrices of regressors and instruments and denote them by $\mathbf X(i)$ and $\mathbf Z(i)$. Thus JIVE removes the dependence between $Z_{i}\hat{\pi}$ and the regressor $X_{i}$.

The corresponding estimate of $\pi$ and the constructed instrument will be:
\begin{equation}
\tilde{\pi}(i)=(\mathbf Z(i)'\mathbf Z(i))^{-1}(\mathbf Z(i)'\mathbf X(i))
\end{equation}

\begin{equation}
Z_{i}\tilde{\pi}(i)=Z_{i}(\mathbf Z(i)'\mathbf Z(i))^{-1}(\mathbf Z(i)'\mathbf X(i)Y)
\end{equation}
$\varepsilon_{i}$ and $X_{j}$ are independent when $i\neq j$, so it follows that
\begin{equation}
E[\varepsilon_{i}Z_{i}\tilde{\pi}(i)]=E[Z_{i}(\mathbf Z(i)'\mathbf Z(i))^{-1}(\mathbf Z(i)')E[\mathbf X(i)\varepsilon_{i}|\mathbf Z]]=0
\end{equation}

Hence, now we have $E[\varepsilon_{i}Z_{i}\tilde{\pi}(i)]=0$, so we have removed the correlation between the fitted values and the structural error presented in equation (4.2).
\par The JIVE estimator is equal to:

\begin{equation}
\hat{\beta}_{JIVE}=(\mathbf {\hat{X}}_{JIVE}'\mathbf X)^{-1}(\mathbf {\hat{X}}_{JIVE}'\mathbf Y)
\end{equation}
where $\hat{X}_{JIVE}$ is $N\times L$ dimensional matrix with the $i$ th row $Z_{i}\tilde{\pi}(i)$. \\

\par \cite{blomquist1999small} summarize the construction of the JIVE estimator by providing the following algorithm:

%\vspace{0.6cm}
\begin{algorithm}
\caption{Jackknife Instrumental Variables Estimator}\label{alg:euclid}
\begin{algorithmic}[1]
\State Use all observations but the $i$th to estimate parameters of the first-stage equation.
\State Combine the estimated first stage parameters with the instruments for the $i$th observation, $Z_i$, to construct a fitted value for the $i$th observation $X_i$.
\State Repeat steps 1 and 2 for all N observations.
\State Regress Y on the fitted values and the exogenous regressors.
\end{algorithmic}
\end{algorithm}

The estimator for $Z_i\pi$, $Z_{i}\tilde{\pi}(i)$ is consistent.
$\hat{\beta}_{JIVE}$ has the same probability limit and first-order asymptotic distribution as $\hat{\beta}_{OPT}$ and $\hat{\beta}_{2SLS}$, under conventional fixed model asymptotics (\cite{stock2002survey}).

\section{Limited Information Maximum-Likelihood Estimator}

Limited Information Maximum Likelihood (LIML) is an alternative method to estimate the parameters of the structural equation. It was formalized by \cite{anderson1949estimation}. The derivation of the LIML estimator is shown in the appendix, from equations (A.13) to (A.19).\\
We follow the same model as described in Section 2, and extend it as shown below: 
\begin{equation}
 \mathbf{Y} = \mathbf{X}\beta + \varepsilon = \mathbf{X_0}\beta_0 + \mathbf{X_1}\beta_1 + \varepsilon
\end{equation}
\begin{equation}
 \mathbf{X_1} = \mathbf{Z}\pi + \eta = \mathbf{Z_0}\pi_0 + \mathbf{Z_1}\pi_1 + \eta
\end{equation}
where $\mathbf{Z_0} = \mathbf{X_0}$.\\
Here, $\mathbf{X_0}$ is the matrix of exogenous variables and $\mathbf{X_1}$ of endogenous variables. $\mathbf{Z_0} = \mathbf{X_0}$ is the matrix of included exogenous variables, $\mathbf{Z_1}$ is the matrix of excluded exogenous variables.
Because the LIML estimator is based on the structural equation for $\mathbf{Y}$ combined with the first-stage equation for $\mathbf{X_1}$, it is called `limited information'. LIML estimator is given by:
\begin{equation}
\hat\beta_{LIML}=(\mathbf X'(\mathbf I-\hat\kappa \mathbf M_Z)\mathbf X)^{-1}\mathbf X'(\mathbf I-\hat\kappa \mathbf M_Z)\mathbf Y
\end{equation}
\par The LIML estimator has some excellent properties when the number of excluded instruments (number of columns in matrix $\mathbf X_1$) in the first-stage equation and the sample size are large. Although the LIML estimator and the 2SLS estimator are asymptotically equivalent in the standard large sample theory, they are quite different in case of many instruments or many weak instruments. It has no finite moments, which implies that its density tends to have very thick tails. \cite{anderson2010asymptotic} show that the LIML estimator shows asymptotic optimality with many weak instruments . The LIML estimator is asmptotically efficient in higher order, while the 2SLS estimator is inconsistent, shown by  \cite{kunitomo1987third}.

\chapter{Simulation Study}
In this chapter, we firstly explore the finite sample behavior of different IV estimators as well as OLS estimator by Monte Carlo simulation. Then, we discuss the performance of all these estimators in 4 models, namely Just-identification with strong IV, Just-identification with weak IV, Over-identification with strong IVs, Over-identification with weak IVs.
The experiment presented here follows from the simulations presented in \cite{angrist1999jackknife}, \cite{davidson2006case}.
\section{Finite Sample Properties}
To maintain simplicity in the design of our experiment, we let the error terms of both reduced-form and structural equation to be homoscedastic, all the regressions to be linear.\\
The structural equation is defined as follow,

\begin{equation}
  Y = \beta_0 \iota + \beta_1 X + \varepsilon
\end{equation}

where $\iota$ is a vector of ones and $X$ is the endogenous variable, which is a one-dimensional column vector generated by the reduced-form equation,

\begin{equation}
  X = Z\pi + \eta = \pi_0 \iota + \displaystyle\sum_{i=1}^{K-1} z_i \pi_i + \eta
\end{equation}

here, the first column of matrix $Z$ is also $\iota$, the remaining columns $z_i$ are IID variables with mean zero and unit variance.\\
Without loss of generality, we set $\beta_0 = \beta _1 = 1$, $\sigma_{\varepsilon}^2 = 1$, $\pi_0 = 0$, $\pi_1,..., \pi_{K-1}$ to be equal. Under the exogeneity condition, $Z$ is uncorrelated with $\varepsilon$, therefore we know that the correlation between $X$
and $\varepsilon$ is only through $\eta$. The correlation coefficient between $\varepsilon$
and $\eta$ is denoted by $\rho$ (the covariance between the two is denoted by $\sigma_{\varepsilon\eta}$).
\par We carry out 3 experiments - we explore how variation in $\rho$, the strength of instruments and also the number of instruments impacts the finite sample behavior of the estimators under study. The limiting $R^2$ of the reduced-form equation, denoted by $R_{\infty}^2$, is given by $R_{\infty}^2 = 1/(1+ \sigma_\eta^2/\parallel \pi \parallel^2)$.
\par We normalized $\sigma_\eta^2 + \parallel \pi \parallel^2$ to 1 so as to restrict $R_{\infty}^2 \in [0, 1]$. $R_{\infty}^2$ is a monotonically increasing function of the `Concentration Parameter', discussed in Section 2. Matrix $Z$ has $K$ columns and $K-2$ over-identifying restrictions.  We vary $\rho$ and $R_{\infty}^2$ at interval of 0.01.
\par Every experiment is performed with sample size 25, 50, 100, 200, 400, 800. All the experiments are replicated 1000 times. Since LIML and JIVE do not have first and second moments, we employ Median Bias, 0.5 quantile of the estimation minus the true value, to evaluate the central tendency of different estimators.\\
In the first experiment, we vary $\rho$, keep $K=7$,  $R_{\infty}^2=0.1$ (implies $\parallel \pi \parallel^2=0.1, \sigma_\eta^2=0.9$) constant. Result is shown in Figure B.1. The median bias of OLS, 2SLS and JIVE proportionally increases  to $\rho$ for all sample sizes. For LIML, as the sample size reaches 200 and above, its median bias is negligible.
\par In the second experiment, we vary $R_{\infty}^2$ and keep $K=7$, $\rho=0.9$ fixed. Figure B.2 clearly shows that the growth of $R_{\infty}^2$ leads to a decline in the median bias for all four estimators. It is noteworthy that at sample sizes of 100 and above, LIML shows much lower median bias (compared to other estimators) even at small values of $R_{\infty}^2$. JIVE shows an interesting trend, it fluctuates around OLS, while maintaining the same overall trend as OLS - decrease in median bias as $R_{\infty}^2$ decreases.
\par We change the value of $K$ in the third experiment and keep $R_{\infty}^2=0.1$, $\rho=0.9$ fixed. In Figure B.3, except for OLS and LIML, increase in number of instruments $K$ leads to growth of median bias. At small sample size (50 or lower), the median bias of LIML increases with $K$, however, at sizes 100 and above, the median bias appears to be stable around zero.

\section{Performance of the Estimators}
We use 4 sets of parameters corresponding to 4 models. The parameter-setting is determined through two hypothesis tests. We employ the first-stage F statistic to detect the strength of instruments and the Anderson-Rubin statistic as a fully robust test for $\beta_1$ in the two cases with weak instruments. We apply the same model in the previous section but fix sample size to be equal to 200 and replicate 5000 times.
To compare estimators, median bias and coverage probability for $95\%$ confidence interval (estimated value plus or minus 1.96 times the asymptotic standard error) are employed. Coverage probability is the proportion of the time that the interval contains the true value, in our case $\beta_1 = 1$. Unlike OLS, 2SLS and LIML, the asymptotic standard error of JIVE is calculated based on $\hat{X}_{JIVE}.$
\par
\begin{itemize}
    \item \textit{Model 1: Just-identification with strong instrument}\\
    Let $\beta_0=\beta_1=1$, $\pi_0=0$, $\pi_1=0.3$. Here, $K=2$ and
    \begin{align*}
        \begin{pmatrix}
        \varepsilon\\
        \eta
        \end{pmatrix} &\sim  N
        \begin{pmatrix}
        \begin{pmatrix}
        0\\
        0
        \end{pmatrix}\!\!,&
        \begin{pmatrix}
        0.25 & 0.20\\
        0.20 & 0.25
        \end{pmatrix}
        \end{pmatrix}
     \end{align*}
         
    \item \textit{Model 2: Just-identification with weak instrument}\\
   Let $\beta_0=\beta_1=1$, $\pi_0=0$, $\pi_1=0.2$. Here, $K=2$ and
   \begin{align*}
        \begin{pmatrix}
        \varepsilon\\
        \eta
        \end{pmatrix} &\sim  N
        \begin{pmatrix}
        \begin{pmatrix}
        0\\
        0
        \end{pmatrix}\!\!,&
        \begin{pmatrix}
        1.0 & 0.9\\
        0.9 & 1.0
        \end{pmatrix}
        \end{pmatrix}
     \end{align*}
             
    \item \textit{Model 3: Over-identification with strong instruments}\\
    Let $\beta_0=\beta_1=1$, $\pi_0=0$, $\pi_1=\pi_2=...=\pi_{15}=0.3$. Here, $K=16$ and
    \begin{align*}
        \begin{pmatrix}
        \varepsilon\\
        \eta
        \end{pmatrix} &\sim  N
        \begin{pmatrix}
        \begin{pmatrix}
        0\\
        0
        \end{pmatrix}\!\!,&
        \begin{pmatrix}
        0.25 & 0.10\\
        0.10 & 0.25
        \end{pmatrix}
        \end{pmatrix}
     \end{align*}
     
\item \textit{Model 4: Over-identification with weak instruments}\\
    Let $\beta_0=\beta_1=1$, $\pi_0=0$, $\pi_1=\pi_2=...=\pi_{15}=0.1$. Here, $K=16$ and
    \begin{align*}
        \begin{pmatrix}
        \varepsilon\\
        \eta
        \end{pmatrix} &\sim  N
        \begin{pmatrix}
        \begin{pmatrix}
        0\\
        0
        \end{pmatrix}\!\!,&
        \begin{pmatrix}
        0.25 & 0.20\\
        0.20 & 0.25
        \end{pmatrix}
        \end{pmatrix}
     \end{align*}
\end{itemize}
Figures B.4 to B.7 present the distribution of $\beta_0$
and $\beta_1$ in Models 1-4 respectively. In Model 1, the OLS estimator of $\beta_1$ is notably biased. In Model 2, the OLS estimator of $\beta_1$ is even more biased. Besides, the 2SLS, LIML and JIVE display a very wide range of estimates for both $\beta_0$ and $\beta_1$. Since we are more interested in the slope coefficient ($\beta_1$), we generate Figures B.8 to B.11 to look at $\beta_1$ in more detail. After we exclude some outliers and focus on the interval $[-3, 5]$, in Figure B.9 we observe notable negative skew in 2SLS, LIML and JIVE estimators.
\par In figure B.6, corresponding to Model 3, we see that $\beta_1$ of the OLS estimator shows a remarkable reduction of bias compared to Model 2. In addition, for all estimators, the distribution of $\beta_1$ is concentrated on a small range which contains the true value. However, in Model 4, the bias of OLS estimate of increases again and the distribution of LIML and JIVE estimates disperses slightly. We include a series of quantiles around $\beta_1$ in Table B.2 to display the dispersion. The JIVE estimator returns a surprisingly large number of outliers on both tails in the just-identified with weak IV model. If we had a smaller number of replications, then in spite of the high coverage probability, we might have seen some extreme JIVE estimates, quite far from the true value.

\par According to the results of our MC simulations, we find that the LIML estimator performs well (in terms of median bias) in most of models, despite the fact that JIVE has the highest coverage probability in all the models. We observe that JIVE does not dominate  LIML in any case/model. Between 2SLS and JIVE, we do not observe JIVE performing uniformly better than 2SLS in the models we considered. While we do observe JIVE do well in overidentified case as discussed in \cite{angrist1999jackknife}, however,
our findings are more in line with results of \cite{davidson2006case}. 

\chapter{Application to Returns to Schooling}

In this section, we revisit the famous paper and its (`provocative' as termed by \cite{bound1995problems}) results that led to the beginning of the weak instruments literature: \cite{angrist1991does}, who use quarter of birth as an instrument for estimating the impact of educational attainment on earnings.
On the same data, we confirm the weakness of the instruments used in the paper using the first-stage F-statistic and proceed to apply estimators more robust to weak instruments.

\par Endogeneity of education is a well-known problem that economists face while estimating the effect of education on earnings. The reason for the endogeneity is omitted variables, such as the `ability', which can be correlated with both educational attainment and earnings of an individual. As a result, OLS will give biased estimates for the return to education. To correct this bias quarter of birth is included in the regression as an instrumental variable.
Association between the quarter of birth and schooling is explained by compulsory schooling requirements in the United States. According to school start age policy, children are required to enter school in the fall of the calendar year in which they turn 6. While compulsory schooling laws allow students to leave school after they turned 16. As a result, students who were born earlier in the calendar year tend to attend school for a shorter period of time than those born at the end of the year. So interaction of the two requirements of schooling laws generates variation in educational attainment for the students who graduate right after their 16th birthday.
\par
Quarter of birth must satisfy the instrument relevance and exogeneity conditions to be a valid instrument for educational attainment. The relevance condition implies that the instrument must be correlated with the endogenous variable, in this case with the years of schooling. Higher the correlation between these two variables, stronger is the first stage in the two-stage least squares estimation. To satisfy the second condition quarter of birth must effect earnings only via its effect on schooling years. In general, it is not possible to test this condition statistically. Angrist and Krueger argue that student's birthday can not be correlated with other personal features which may affect earnings, thus the variation in education due to the individual's birthday is exogenous (\cite{angrist1991does}).
\par For our estimations, we used the dataset from \cite{angrist1991does} which is taken from 1980 US Census. The sample consists of 329,509 men, who were born in 1930-1939. The dataset includes information on the quarter of birth, year of birth, state of birth, years of schooling and earnings for this sample. All figures and tables are presented in Appendix B.\\
Figure B.12 shows the relation between quarter of birth and schooling (first-stage). The graph indicates that there is an upward trend in average years of schooling for men born in 1930-1939. There is also a persistent seasonal pattern in education. Men born at the beginning of the year tend to have less schooling on average than those who were born later in the year.
\par Figure B.13 illustrates the reduced form, which is the relationship between the quarter of birth and wages. Here we also notice a pattern where the 3rd and 4th quarter of births correspond to a higher log weekly wages.
\par
Our general model is given by:
\begin{equation}
E_{i}=\sum_{c}Y_{ic}\delta _{c}+\sum_{c}\sum_{j}Y_{ic}Q_{ij}\Theta _{jc}+\epsilon _{i}
\end{equation}
\begin{equation}
lnW_{i}=\sum_{c}Y_{ic}\xi _{c}+\rho E_{i}+\mu _{i}
\end{equation}

Here $E_{i}$ is the schooling of\textit{ i}th individual, $Y_{ic}$ is a dummy variable, denoting if the \textit{i}th individual was born in \textit{c}th year, $Q_{ij}$ is a dummy variable showing quarter of birth of the \textit{i}th individual and $W_{i}$ is the weakly wage.
In Table B.3 we present results of the first stage estimations. The first and third columns of the table indicate that individuals born in the last quarter of the year had about 0.10 year more schooling compared to the men born in the first to third quarters. The second, fourth and fifth columns show the estimates of each quarter of birth in comparison with the first quarter. Naturally, the largest difference is between the last and the first quarter, which is around 0.15 year, independent of including year of birth and state of birth as control variables in the regression.
\par
Table B.4 attempts to replicate the main results of Table 2 from \cite{angrist1999jackknife}.
We compare 2SLS, LIML and JIVE estimators for three different specifications of the model based on \cite{angrist1991does} data. Additionally, we report the first-stage F-statistic and $R^2$ values.
\par Column 1 shows the results of our estimation wherein we regress wages on schooling, including three quarter of birth dummies as instruments and nine year of birth dummies as control variables. Here, our first stage F-statistic is greater than the critical value (for the case of K = 3 where K is the number of instruments) of 9.08, according to Table B.1 provided in Appendix B. Hence in the first model the excluded instruments are strong.
\par
Moving to column 2, to control for the age-related trends we include interactions of the year of birth dummies with quarter of birth dummies as instruments in the second model. This leads to a slightly lower 2SLS estimate as well as a slight decrease in standard errors. However, the value of the first-stage F-statistic reduces dramatically to 4.91. The small value of the F-statistic indicates that the excluded instruments and educational attainment are only weakly correlated, which as we discussed in section 2.4.2 can lead to finite-sample bias in the 2SLS estimate. Looking at the $R^{2}$ in columns (1) and (2), we notice that compared to the case of the first column, the explanatory power of the instruments does not increase much in the second specification.
\par
Finally, in the last specification (column 3), we increase the number of instruments to 180, taking year of birth $\times$ quarter of birth, and state of birth $\times$ quarter of birth interactions as instruments. The reasoning behind increasing the number of instruments by including interaction terms is to ensure that seasonal differences do not vary by state and birth year. While this results in about 40 percent reduction in standard errors compared to the column (2), however, the value of the first-stage F-statistic again decreases, compared to the preceding model. This indicates that while attempting to increase the precision of the 2SLS estimates, the weakness of the instruments, however, leads to the 2SLS estimates becoming biased.
\par

In Table B.4 we show the estimates of the alternative estimators discussed previously (LIML and JIVE) which are considered to have better finite sample properties when the instruments are weak. As we can see, JIVE estimates vary considerably from the other two.
\par
Given the encouraging performance of the LIML estimator in our simulation study, and considering the closeness of the LIML and 2SLS estimates in our application to returns to schooling, we suspect that in this specific case, LIML and 2SLS probably give more reliable results than JIVE. However, we do not at all recommend using 2SLS estimator in case of weak instruments in general, considering the theoretical discussion and simulation results presented previously.

\chapter{Conclusion}
In our study, we hope to have provided the reader a concise but comprehensive overview of selected  aspects of the weak instruments literature.\\
When we face the issue of weak instruments, the best solution is of course to find better, stronger instruments. However, in empirical practice this is easier said than done. Hence, we think research into alternative estimators which can give more reliable estimates than the two-stage least squares estimator in the weak instruments case, is very relevant. From our small study, we find the LIML estimator to perform the best when the correlation between the instrument and the endogenous explanatory variable is low. We posit the LIML be a possible solution in the (fairly common) case that a researcher has a weak instrument(s) and is not in a position to find other (stronger) instruments. Development of methods for robust inference of weak instruments such as the Anderson-Rubin statistic is a very promising area of further research.

\appendix
\chapter{Formulae}
\pagenumbering{roman}
\section{Deriving the approximate expression of 2SLS Bias}
This derivation has been adopted from \cite{results}. Start with the representation of the 2SLS estimator as shown in equation (2.8).
\begin{equation}
\hat{\beta }_{2SLS}=(\textbf{X}'P_{Z}\textbf{X})^{-1}\textbf{X}'P_{Z}\textbf{Y}=\beta+(\textbf{X}'P_{Z}\textbf{X})^{-1}\textbf{X}'P_{Z}\varepsilon
\end{equation}
where $P_{Z}=\textbf{Z}(\textbf{Z}'\textbf{Z})^{-1}\textbf{Z}'$ is the projection matrix.
So the bias of $\hat{\beta }_{2SLS}$ will be
\begin{equation}
\hat{\beta }_{2SLS}-\beta = (\textbf{X}'P_{Z}\textbf{X})^{-1}(\pi'\textbf{Z}'+\eta')P_{Z}\varepsilon
                          = (\textbf{X}'P_{Z}\textbf{X})^{-1}\pi'\textbf{Z}'\varepsilon +(\textbf{X}'P_{Z}\textbf{X})^{-1}\eta' P_{Z}\varepsilon
\end{equation}
Using group asymptotics the expectation of this expression can presented as:
\begin{equation}
    E[\hat{\beta }_{2SLS}-\beta ]\approx (E[{\textbf{X}}'P_{Z}\textbf{X}])^{-1}E[{\pi}'{\textbf{Z}}'\varepsilon ]+(E[{\textbf{X}}'P_{Z}\textbf{X}])^{-1}E[{\eta }'P_{Z}\varepsilon ]
\end{equation}
$Z_{i}$ instruments are uncorrelated with $\varepsilon_{i}$ and $\eta_{i}$, so $E[{\pi }'{\textbf{Z}}'\varepsilon ]=0 $ and we will have
\begin{equation}
    E[\hat{\beta }_{2SLS}-\beta ]\approx (E[{\textbf{X}}'P_{Z}\textbf{X}])^{-1}E[{\pi}'{\textbf{Z}}'\varepsilon ]+(E[{\textbf{X}}'P_{Z}\textbf{X}])^{-1}E[{\eta }'P_{Z}\varepsilon ]=(E[{\textbf{X}}'P_{Z}\textbf{X}])^{-1}E[{\eta }'P_{Z}\varepsilon]
\end{equation}
Substituting the first stage equation $\textbf{X}=\textbf{Z}\pi +\eta$ we have
\begin{equation}
E[\hat{\beta }_{2SLS}-\beta ]\approx (E[({\pi }'{\textbf{Z}}'+{\eta }')P_{Z}(\textbf{Z}\pi +\eta )])^{-1}E[{\eta }'P_{Z}\varepsilon ]
\end{equation}
We have that $E[{\pi }'{\textbf{Z}}'\eta ]=0 $, so
\begin{equation}
    E[\hat{\beta }_{2SLS}-\beta ]\approx[E({\pi }'{\textbf{Z}}'\textbf{Z}\pi )+E({\eta }'P_{Z}\eta )]^{-1}E({\eta }'P_{Z}\varepsilon )
\end{equation}
Notice that ${\eta }'P_{Z}\eta$ is a scalar and is equal to its trace. $P_{Z}$ is an idempotent matrix, so its trace is equal to its rank, Q. So
\begin{equation}
    E({\eta }'P_{Z}\eta)=E[tr({\eta }'P_{Z}\eta)]=E[tr(P_{Z}\eta{\eta }')]=tr(P_{Z}E[\eta{\eta }'])=tr(P_{Z}\sigma _{\eta }^{2}I)=\sigma _{\eta }^{2}tr(P_{Z})=\sigma _{\eta }^{2}Q
\end{equation}
With a similar technique we can show that  $E({\eta }'P_{Z}\varepsilon)$ is equal to $\sigma _{\varepsilon \eta }Q$.
Substituting these results in equation (A.6) we have
\begin{equation}
   E[\hat{\beta }_{2SLS}-\beta]\approx\sigma _{\varepsilon \eta }Q[E({\pi }'{\textbf{Z}}'\textbf{Z}\pi )+\sigma _{\eta }^{2}Q]^{-1}=\frac{\sigma_{\varepsilon\eta }}{\sigma  _{\eta }^{2}}[\frac{E(\pi'\textbf{Z}'\textbf{Z}\pi)/Q}{\sigma _{\eta }^{2}}+1]^{-1}
\end{equation}
The population F-statistic for the first stage regression is the following
\begin{equation}
    F=\frac{E(\pi'\textbf{Z}'\textbf{Z}\pi)/Q}{\sigma _{\eta }^{2}}
    \end{equation}
So (A.8) can be expressed as
\begin{equation}
    E[\hat{\beta }_{2SLS}-\beta]\approx \frac{\sigma_{\varepsilon\eta }}{\sigma  _{\eta }^{2}}\frac{1}{F+1}
\end{equation}

Assume that the ${\pi}$ coefficients are zero and $F=0$. In this case $\sigma_{X}^{2}=\sigma_{\eta }^{2}$ and
\begin{equation}
    E[\hat{\beta }_{2SLS}-\beta]\approx \frac{\sigma_{\varepsilon\eta }}{\sigma  _{X}^{2}}
\end{equation}

Thus, when $\pi\neq 0$ and F is small, then 2SLS will be biased towards OLS. 

Note that $\frac{\sigma_{\varepsilon\eta }}{\sigma  _{X}^{2}}$ is also the bias of OLS estimator, because when $\pi=0$, $cov(\varepsilon _{i},X_{i})=\sigma _{\eta \varepsilon }$ and
\begin{equation}
    \beta _{OLS} = \frac{cov(Y_{i},X_{i})}{var(X_{i})} = \frac{cov(\beta X_{i}+\varepsilon  _{i},X_{i})}{var(X_{i})} = \beta + \frac{cov(\varepsilon  _{i},X_{i})}{var(X_{i})} = \frac{\sigma _{\varepsilon \eta }}{\sigma _{X}^{2}}
\end{equation}
Hence in this case OLS and 2SLS estimators on average are the same. And if $\pi$ is different from zero, $\hat{\beta }_{2SLS}$ will be biased in the direction of OLS estimator. If we add weak instruments to the regression bias of 2SLS will only increase. \\

\newpage

\section{Deriving the LIML Estimator}
LIML Estimator Derivation\cite{brucehansen2019}\cite{davidson2003econometric}:\\
Structural Equation:
\begin{equation}
 \mathbf{Y} = \mathbf{X}\beta + \varepsilon = \mathbf{X_0} \beta_0 + \mathbf{X_1} \beta_1 + \varepsilon
\end{equation}
Reduced-form Equation:
\begin{equation}
  \mathbf{X_1} = \mathbf{Z}\pi + \eta = \mathbf{Z_0} \pi_0 + \mathbf{Z_1} \pi_1 + \eta
\end{equation}
Here, $\mathbf{X_0}$ is the matrix of exogenous variables and $\mathbf{X_1}$ of endogenous variables. $\mathbf{Z_0} = \mathbf{X_0}$ is the matrix of included exogenous variables, $\mathbf{Z_1}$ is the matrix of excluded exogenous variables. \\
The LIML estimate of $\beta_1$ in (A.13) is given by minimize the ratio:
\begin{equation}
 \kappa \equiv \frac{(\mathbf{Y}-\mathbf{X_1}\beta_1)'\mathbf{M_{X_0}}(\mathbf{Y}-\mathbf{X_1}\beta_1)}{(\mathbf{Y}-\mathbf{X_1}\beta_1)'\mathbf{M_Z}(\mathbf{Y}-\mathbf{X_1}\beta_1)} \equiv \frac{\mathbf{\gamma'}\mathbf{Y'_\ast} \mathbf{M_{X_0}}\mathbf{Y_\ast} \mathbf{\gamma}}{\mathbf{\gamma'}\mathbf{Y'_\ast} \mathbf{M_Z} \mathbf{Y_\ast} \mathbf{\gamma}}
\end{equation}
where $\mathbf{Y_\ast} \equiv [\mathbf{Y}\;\; \mathbf{X_1}]$, \;\;$\mathbf{\gamma}=[1 \;\vdots -\beta_1]$, \;\;$\mathbf{M_Z} = \mathbf{I} - \mathbf{Z}(\mathbf{Z'}\mathbf{Z})^{-1}\mathbf{Z'}$, \;\; $\mathbf{M_{X_0}} = \mathbf{I} - \mathbf{X_0}(\mathbf{X_0'}\mathbf{X_0})^{-1}\mathbf{X_0'}$.\\
The first order conditions obtained by differentiating the rightmost expression in (A.15) with respect to $\gamma$ are
\begin{equation}
 2\mathbf{Y'_\ast} \mathbf{M_{X_0}}\mathbf{Y_\ast} \mathbf{\gamma}(\mathbf{\gamma'}\mathbf{Y'_\ast} \mathbf{M_Z} \mathbf{Y_\ast} \mathbf{\gamma}) -  2\mathbf{Y'_\ast} \mathbf{M_Z} \mathbf{Y_\ast} \mathbf{\gamma}(\mathbf{\gamma'}\mathbf{Y'_\ast} \mathbf{M_{X_0}} \mathbf{Y_\ast} \mathbf{\gamma}) = \mathbf{0}
\end{equation}
Premultiplying (A.4) by $(\mathbf{Y'_\ast} \mathbf{M_Z} \mathbf{Y_\ast})^{-1/2}$ and inserting that factor multiplied by its inverse before $\gamma$, we yield
\begin{equation}
 ((\mathbf{Y'_\ast} \mathbf{M_Z} \mathbf{Y_\ast})^{-1/2} \mathbf{Y'_\ast} \mathbf{M_{X_0}} \mathbf{Y_\ast} (\mathbf{Y'_\ast} \mathbf{M_Z} \mathbf{Y_\ast})^{-1/2} - \kappa \mathbf{I})\gamma_\ast = \mathbf{0}
\end{equation}
where $\mathbf{\gamma_\ast} \equiv (\mathbf{Y'_\ast} \mathbf{M_Z} \mathbf{Y_\ast})^{1/2}\mathbf{\gamma}$ \;and $\mathbf{\gamma}$ is an eigenvalue of $(\mathbf{Y'_\ast} \mathbf{M_Z} \mathbf{Y_\ast})^{-1/2} \mathbf{Y'_\ast} \mathbf{M_{X_0}} \mathbf{Y_\ast} (\mathbf{Y'_\ast} \mathbf{M_Z} \mathbf{Y_\ast})^{-1/2}$.\\
$\hat{\kappa}$ is the smallest eigenvalue, because $\hat{\kappa}$ is the minimum value of (A.15).\\
The LIML estimator of $\beta$ is defined by estimating equation
\begin{equation}
 \mathbf{X}'(\mathbf{I}-\hat{\kappa}\mathbf{M_Z})(\mathbf{Y}-\mathbf{X}\hat{\beta}_{LIML}) = \mathbf{0}
\end{equation}
Once $\hat{\kappa}$ has been computed, we find that
\begin{equation}
 \hat{\beta}_{LIML} = (\mathbf{X'}(\mathbf{I}-\hat{\kappa}\mathbf{M_Z})\mathbf{X})^{-1} \mathbf{X'}(\mathbf{I}-\hat{\kappa}\mathbf{M_Z})\mathbf{Y}
\end{equation}

% Tables
%%%%%%%%%%%%%%%%%%%%%%%%%%%%%%%%%%%%%%%%%%%%%%%%%%%%%%
\chapter{Tables \& Figures}
\begin{table*}[t]
\centering
\caption{\;Selected Critical Values for Weak Instrument Tests for TSLS Based on the First-stage F statistic. Source:\cite{stock2002testing}, table 1.}
\label{tab1}
\begin{tabular}{c c r}
    \toprule
    \midrule
    \multirow{1}[8]{*}{number of IV (K)} &
    \multicolumn{2}{c}{Relative bias $>10\%$}\\
    \cmidrule(rl){2-3}& Threshold $\mu^2/K$  & F statistic critical value $5\%$ \\
    \cmidrule(r){1-1}\cmidrule(l){2-3}\\
    \multicolumn{1}{l}{3}& 3.71 & 9.08\\
    \multicolumn{1}{l}{5}& 5.82 & 10.83\\
    \multicolumn{1}{l}{10}& 7.41 & 11.49\\
    \multicolumn{1}{l}{15}& 7.94 & 11.51\\
    \midrule
    \bottomrule
\end{tabular}
\end{table*}

\begin{table*}[t]
\centering
\caption{\;Performance of the estimators under Just-identification with strong IV, Just-identification with weak IV, Over-identification with strong IVs, Over-identification with weak IVs}
\label{tab2}
\begin{tabular}{c r r r r r r r}
    \toprule
    \midrule
    \multirow{2}[4]{*}{Model  1} & \multicolumn{5}{c}{Quantiles around $\beta_1$} & & \multirow{1}[4]{*}{Coverage Prob.}\\ 
    \cmidrule(rl){2-6}& $0\%$  & $25\%$ & $50\%$ & $75\%$ & $100\%$  
    & Median Bias & {$C.I._{95\%}$}\\
    \cmidrule(r){1-1}\cmidrule(l){2-8}\\
    \multicolumn{1}{l}{OLS}& 0.43 & 0.56 & 0.59 & 0.62 & 0.75 & 0.587 & 0.000\\
    \multicolumn{1}{l}{2SLS}& -0.58 & -0.08 & 0.00 & 0.07 & 0.37 & 0.000 & 0.536\\
    \multicolumn{1}{l}{LIML}& -1.39 & -0.09 & 0.00 & 0.07 & 0.31 & 0.000 & 0.539\\
    \multicolumn{1}{l}{JIVE}& -0.74 & -0.11 & -0.02 & 0.06 & 0.36 & -0.023 & 0.999\\
    \midrule
    \multirow{2}[4]{*}{Model  2} & \multicolumn{5}{c}{Quantiles around $\beta_1$}& & \multirow{1}[4]{*}{Coverage Prob.}\\ 
    \cmidrule(rl){2-6}& $0\%$  & $25\%$ & $50\%$ & $75\%$ & $100\%$  
    & Median Bias & {$C.I._{95\%}$}\\
    \cmidrule(r){1-1}\cmidrule(l){2-8}\\
    \multicolumn{1}{l}{OLS}& 0.74 & 0.84 & 0.86 & 0.89 & 0.98 & 0.864 & 0.000\\
    \multicolumn{1}{l}{2SLS}& -242.30 & -0.30 & 0.00 & 0.20 & 80.93 & 0.000 & 0.141\\
    \multicolumn{1}{l}{LIML}& -404.49 & -0.23 & 0.06 & 0.21 & 1531.56 & 0.059 & 0.145\\
    \multicolumn{1}{l}{JIVE}& -628.81 & -0.74 & -0.16 & 0.15 & 2182.60 & -0.160 & 0.846\\
    \midrule
    \multirow{2}[4]{*}{Model  3} & \multicolumn{5}{c}{Quantiles around $\beta_1$}& & \multirow{1}[4]{*}{Coverage Prob.}\\ 
    \cmidrule(rl){2-6}& $0\%$  & $25\%$ & $50\%$ & $75\%$ & $100\%$  
    & Median Bias & {$C.I._{95\%}$}\\
    \cmidrule(r){1-1}\cmidrule(l){2-8}\\
    \multicolumn{1}{l}{OLS}& -0.05 & 0.04 & 0.06 & 0.08 & 0.16 & 0.063 & 0.370\\
    \multicolumn{1}{l}{2SLS}& -0.10 & -0.01 & 0.00 & 0.03 & 0.11 & 0.005 & 0.925\\
    \multicolumn{1}{l}{LIML}& -0.11 & -0.02 & 0.00 & 0.02 & 0.11 & 0.000 & 0.921\\
    \multicolumn{1}{l}{JIVE}& -0.11 & -0.02 & 0.00 & 0.02 & 0.11 & 0.000 & 0.999\\
    \midrule
    \multirow{2}[4]{*}{Model  4} & \multicolumn{5}{c}{Quantiles around $\beta_1$}& & \multirow{1}[4]{*}{Coverage Prob.}\\ 
    \cmidrule(rl){2-6}& $0\%$  & $25\%$ & $50\%$ & $75\%$ & $100\%$  
    & Median Bias & {$C.I._{95\%}$}\\
    \cmidrule(r){1-1}\cmidrule(l){2-8}\\
    \multicolumn{1}{l}{OLS}& 0.33 & 0.47 & 0.50 & 0.53 & 0.66 & 0.500 & 0.000\\
    \multicolumn{1}{l}{2SLS}& -0.24 & 0.03 & 0.08 & 0.14 & 0.32 & 0.085 & 0.477\\
    \multicolumn{1}{l}{LIML}& -0.61 & -0.07 & 0.00 & 0.02 & 0.11 & 0.000 & 0.921\\
    \multicolumn{1}{l}{JIVE}& -0.61 & -0.08 & -0.01 & 0.05 & 0.28 & -0.014 & 0.996\\
    \midrule
    \bottomrule
\end{tabular}
\end{table*}

% Tables for real data application
%%%%%%%%%%%%%%%%%%%%%%%%%%%%%%%%%%%%%%%%%%%
\begin{table}[]
\centering
\caption{ First stage regression results(dependent variable: Years of schooling}
\label{tab}
\begin{tabular}{lccccc}

 \\ \hline
Regressor & (1) & (2) & (3) & (4) & (5) \\ \hline
quarter 2 &  & \begin{tabular}[c]{@{}c@{}}0.057\\ (0.016)\end{tabular} &  & \begin{tabular}[c]{@{}c@{}}0.056\\ (0.016)\end{tabular} & \begin{tabular}[c]{@{}c@{}}0.045\\ (0.016)\end{tabular} \\
quarter 3 &  & \begin{tabular}[c]{@{}c@{}}0.117\\ (0.016)\end{tabular} &  & \begin{tabular}[c]{@{}c@{}}0.113\\ (0.016)\end{tabular} & \begin{tabular}[c]{@{}c@{}}0.11\\ (0.016)\end{tabular} \\
quarter 4 & \begin{tabular}[c]{@{}c@{}}0.092\\ (0.013)\end{tabular} & \begin{tabular}[c]{@{}c@{}}0.151\\ (0.016)\end{tabular} & \begin{tabular}[c]{@{}c@{}}0.09\\ (0.013)\end{tabular} & \begin{tabular}[c]{@{}c@{}}0.148\\ (0.016)\end{tabular} & \begin{tabular}[c]{@{}c@{}}0.156\\ (0.038)\end{tabular} \\ \hline
9 year of birth dummies &  &  & \checkmark & \checkmark & \checkmark \\ \hline
50 state of birth dummies &  &  &  &  & \checkmark \\ \hline
\end{tabular}
\end{table}

\begin{table}[]
\centering
\caption{ 2SLS, LIML and JIVE estimates of the economic returns to schooling}
\label{tab:table 1}
\begin{tabular}{lccc}
\hline
\\ \hline
 & (1) & (2) & (3) \\
2SLS & \begin{tabular}[c]{@{}c@{}}0.105\\ (0.020)\end{tabular} & \begin{tabular}[c]{@{}c@{}}0.089\\  (0.016)\end{tabular} & \begin{tabular}[c]{@{}c@{}}0.093\\  (0.009)\end{tabular} \\
LIML & \begin{tabular}[c]{@{}c@{}}0.106\\ (0.020)\end{tabular} & \begin{tabular}[c]{@{}c@{}}0.093\\  (0.018)\end{tabular} & \begin{tabular}[c]{@{}c@{}}0.106\\  (0.012)\end{tabular} \\
JIVE & \begin{tabular}[c]{@{}c@{}}0.467\\  (0.001)\end{tabular} & \begin{tabular}[c]{@{}c@{}}0.472\\  (0.001)\end{tabular} & \begin{tabular}[c]{@{}c@{}}0.499\\  (0.003)\end{tabular} \\
F-statistic (first stage) & 32.27 & 4.91 & 2.58 \\ \hline
$R^{2}$(first stage, $\times$ 100) & 0.029 & 0.044 & 0.14 \\ \hline
Adjusted $R^{2}$(first stage, $\times$ 100) & 0.028 & 0.036 & 0.086 \\ \hline
\textit{Controls} &  &  &  \\ \hline
Year of birth & \checkmark & \checkmark & \checkmark \\
State of birth &  &  & \checkmark \\ \hline
\textit{Excluded instruments} &  &  &  \\ \hline
Quarter-of-birth dummies & \checkmark &  &  \\
Quarter of birth*year of birth &  & \checkmark & \checkmark \\
Quarter of birth*state of birth &  &  & \checkmark \\ \hline
Number of instruments & 3 & 30 & 180 \\ \hline
\end{tabular}
\end{table}

% Figures
%%%%%%%%%%%%%%%%%%%%%%%%%%%%%%%%%%%%%%%%%%%%%%%%%%%%%%%
\newpage
\begin{figure}
\centering
\caption{$\:R_{\infty}^2 = 0.1,\: K = 7$ }
\includegraphics[width=\textwidth]{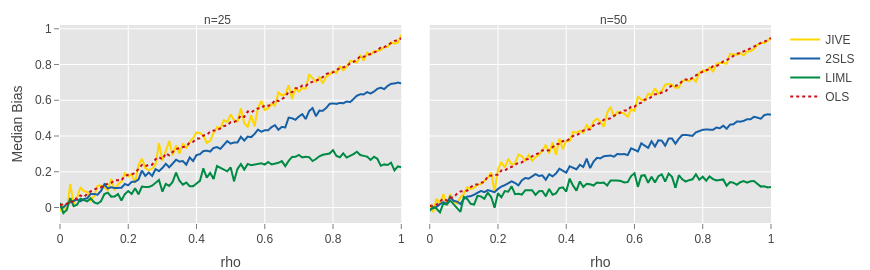}\\
\includegraphics[width=\textwidth]{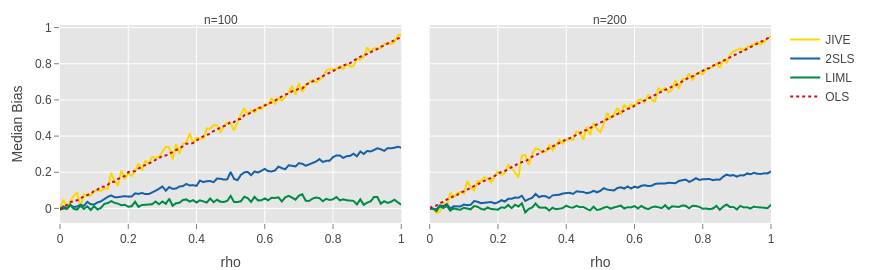}\\
\includegraphics[width=\textwidth]{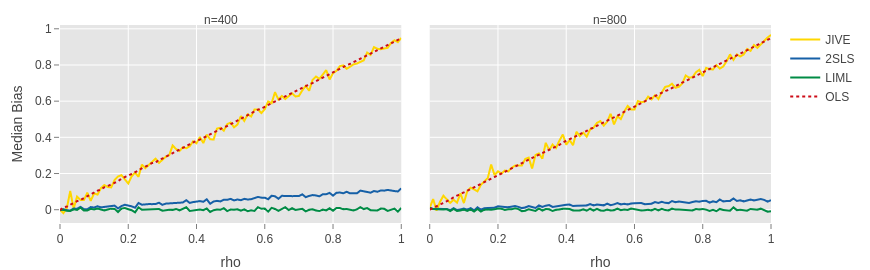}\\
\end{figure}

\begin{figure}
\centering
\caption{$\:\rho = 0.9,\: K = 7$ }
\includegraphics[width=\textwidth]{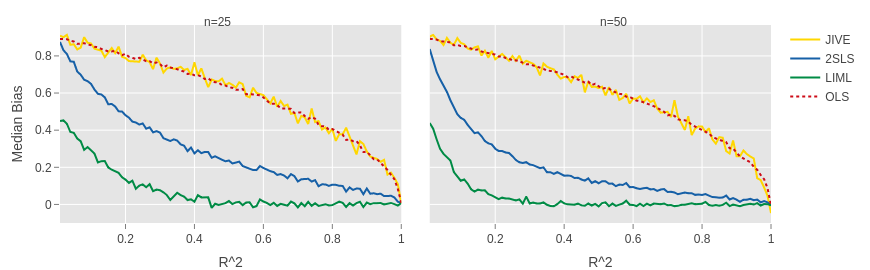}\\
\includegraphics[width=\textwidth]{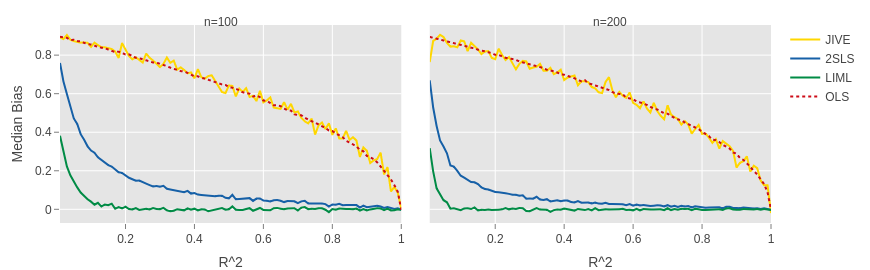}\\
\includegraphics[width=\textwidth]{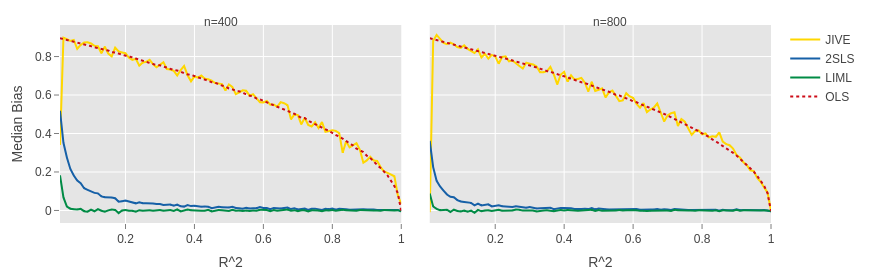}\\
\end{figure}

\begin{figure}
\centering
\caption{$\:R_{\infty}^2 = 0.1,\:\rho = 0.9$ }
\includegraphics[width=\textwidth]{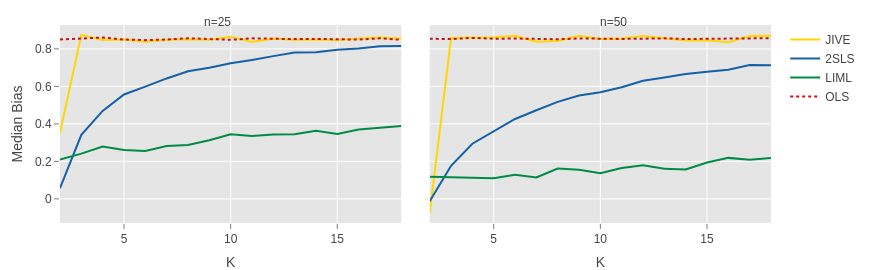}\\
\includegraphics[width=\textwidth]{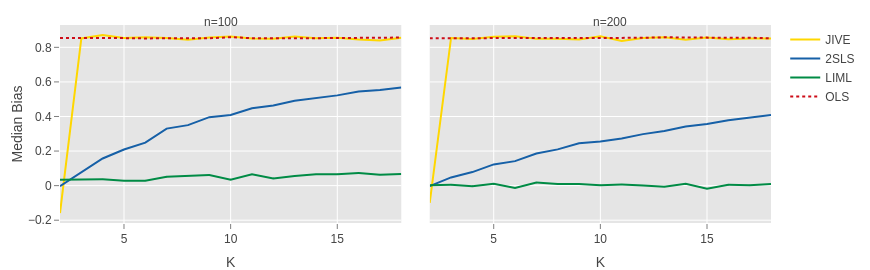}\\
\includegraphics[width=\textwidth]{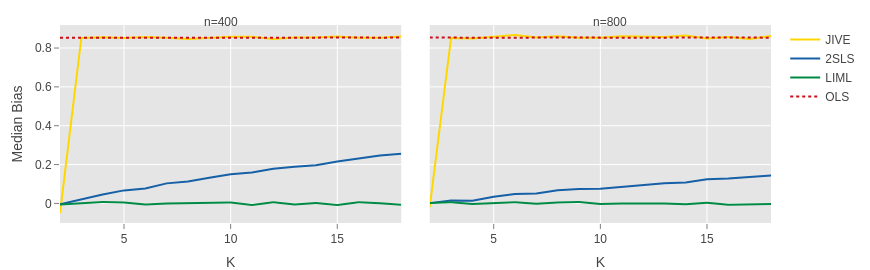}\\
\end{figure}
%%%%%%%%%%%%%%%%%%%%%%%%%%%%%%%%%%%%%%%%%%%%%%%%%%%%%%%
\newpage
\begin{figure}[h]
\caption{\;Joint distribution of $\beta$ in Model 1.}
\centering
\includegraphics[width=0.6\textwidth]{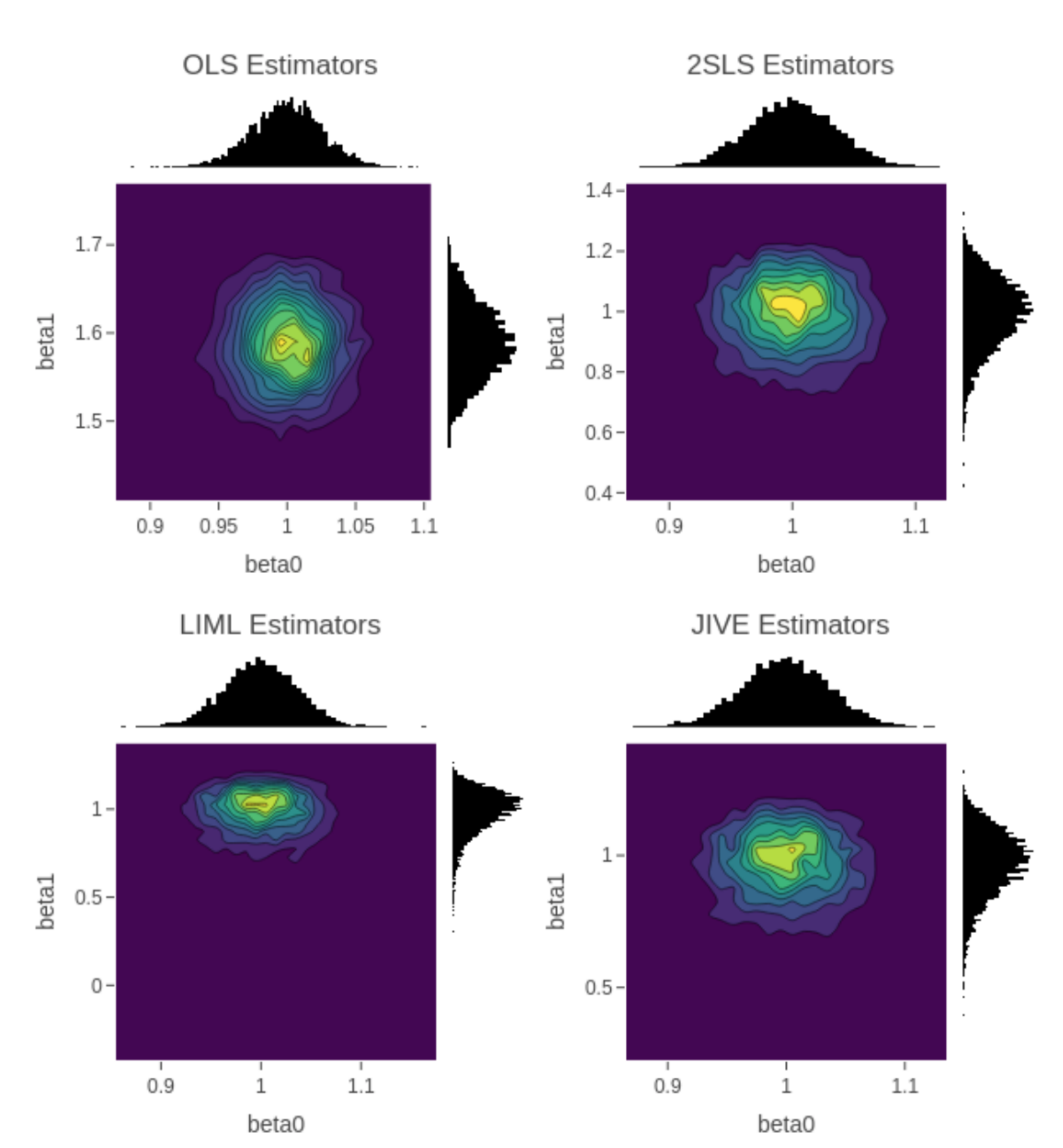}
\end{figure}

\begin{figure}[h]
\caption{\;Joint distribution of $\beta$ in Model 2.}
\centering
\includegraphics[width=0.6\textwidth]{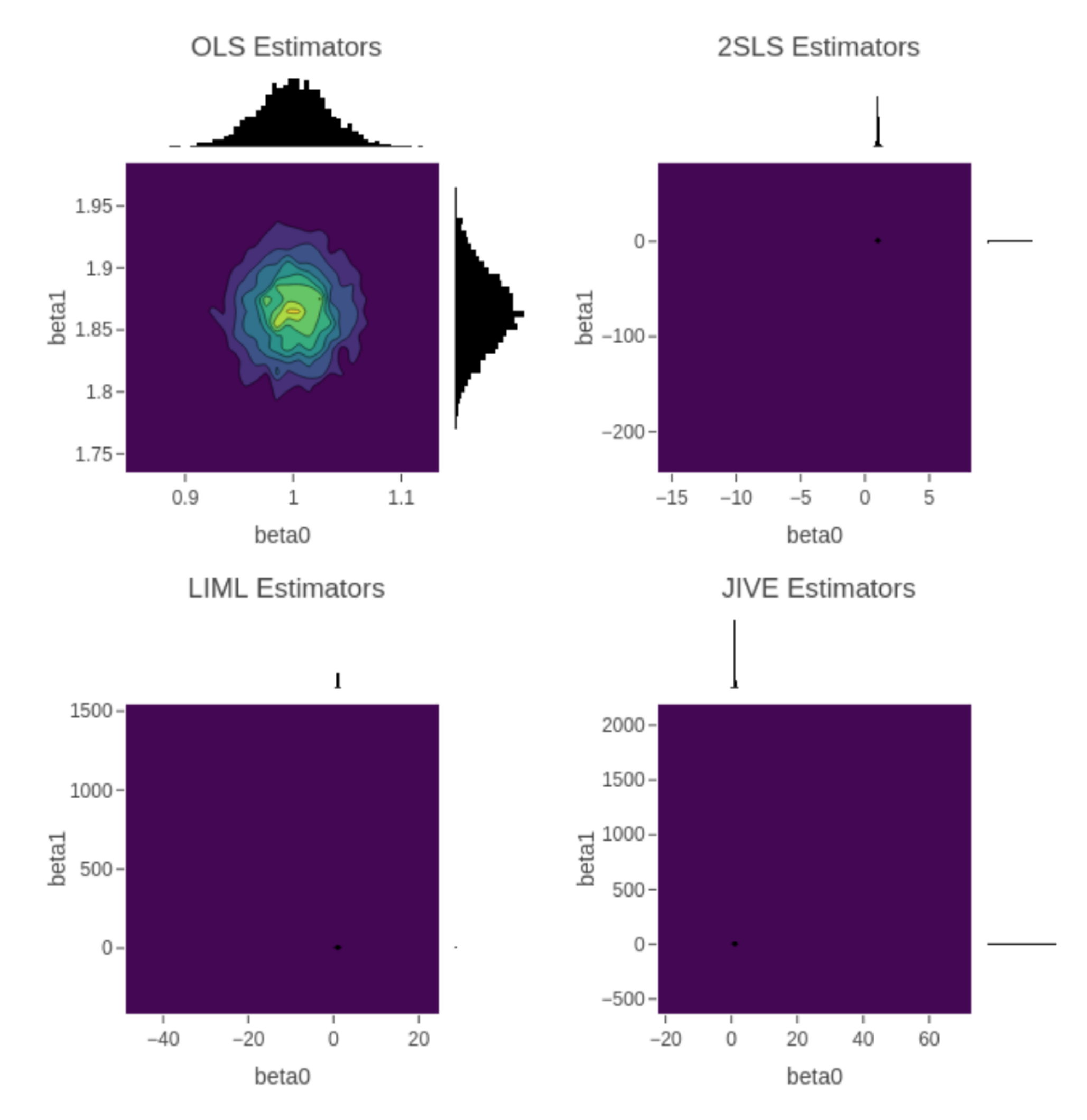}
\end{figure}

\begin{figure}[h]
\caption{\;Joint distribution of $\beta$ in Model 3.}
\centering
\includegraphics[width=0.6\textwidth]{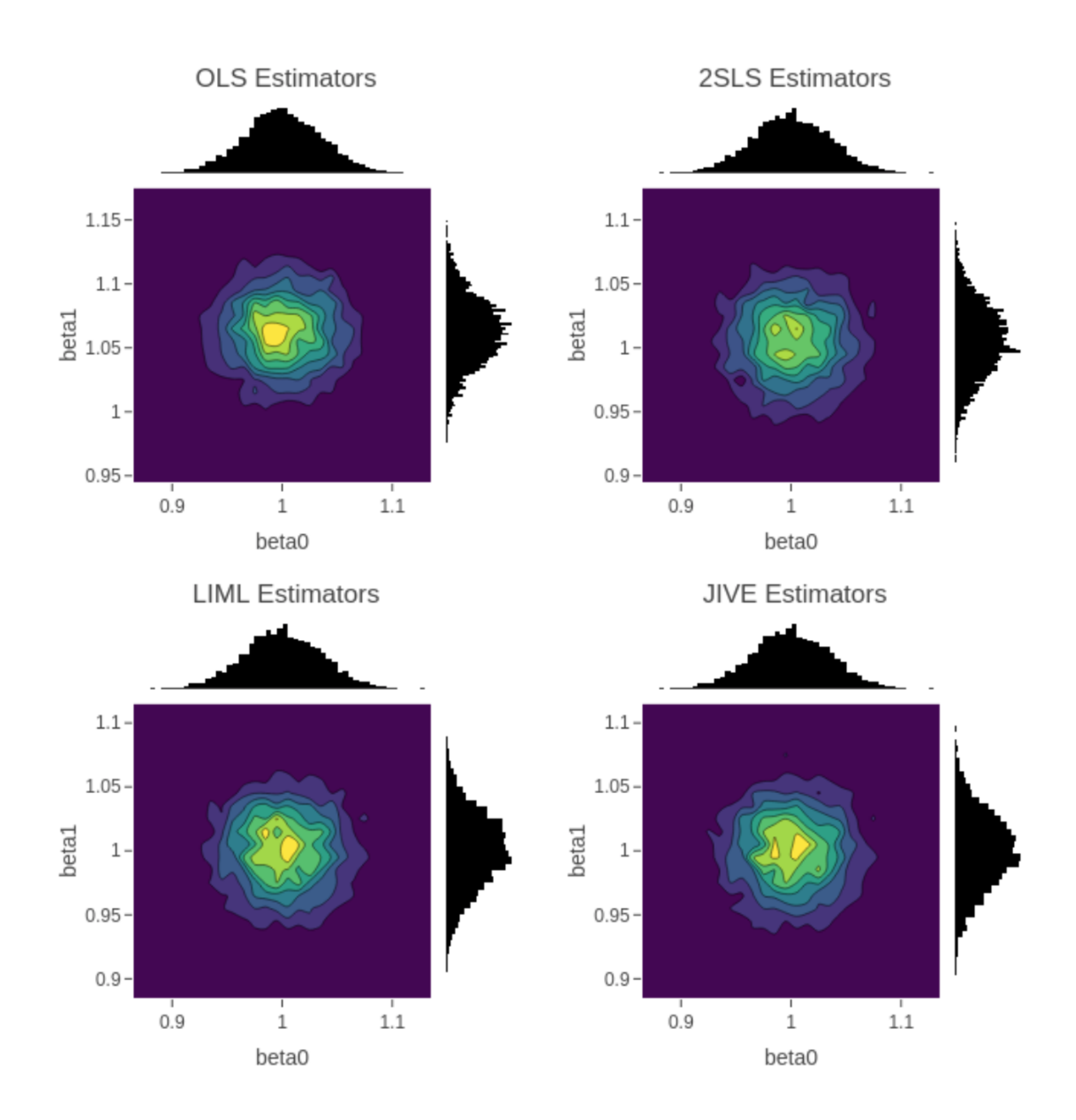}
\end{figure}

\begin{figure}[h]
\caption{\;Joint distribution of $\beta$ in Model 4.}
\centering
\includegraphics[width=0.6\textwidth]{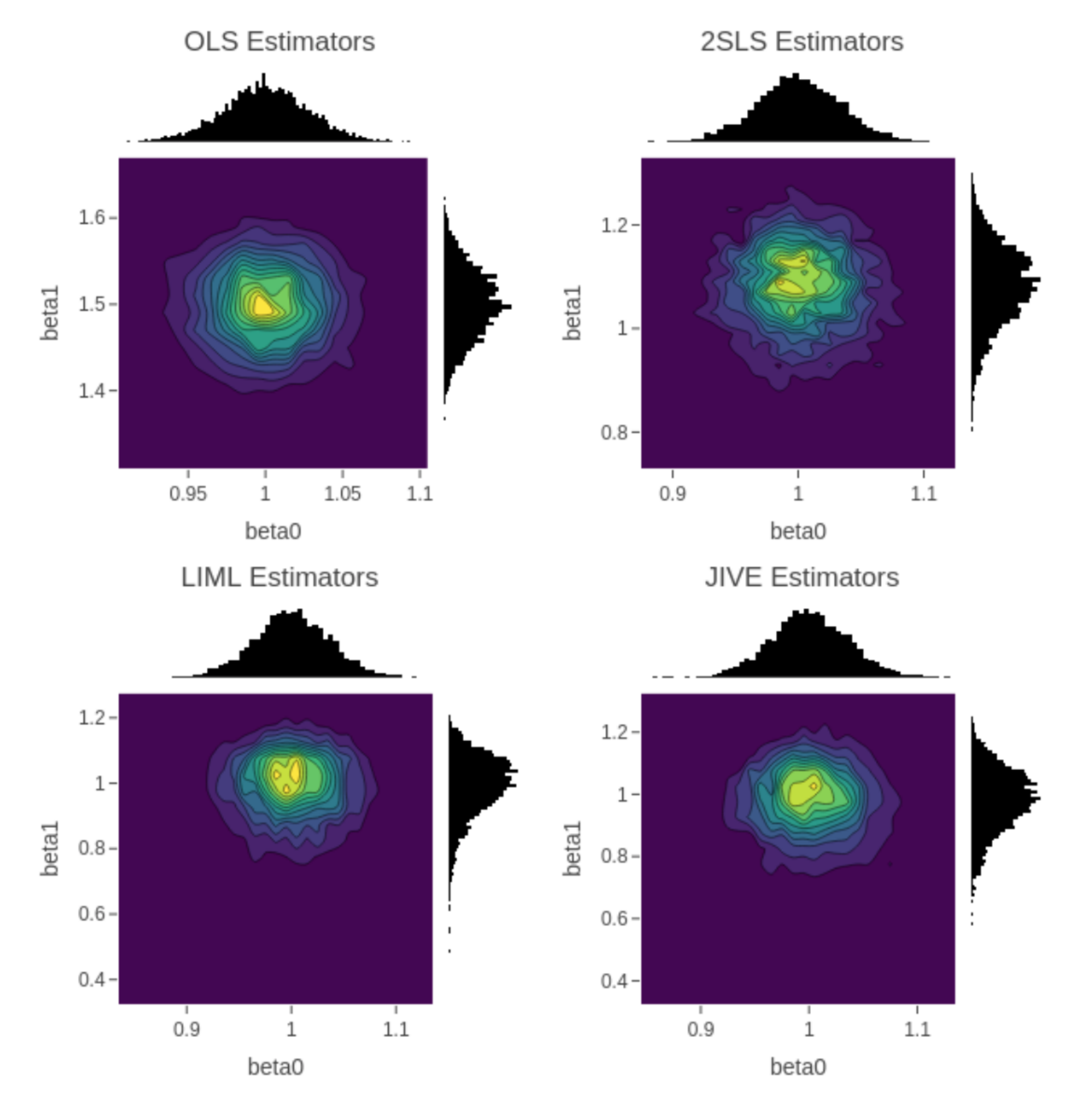}
\end{figure}
%%%%%%%%%%%%%%%%%%%%%%%%%%%%%%%%%%%%%%%%%%%%%%%%%%%%%%
\begin{figure}[h]
\caption{\;Distribution of $\beta_1$ in Model 1.}
\centering
\includegraphics[width=0.9\textwidth]{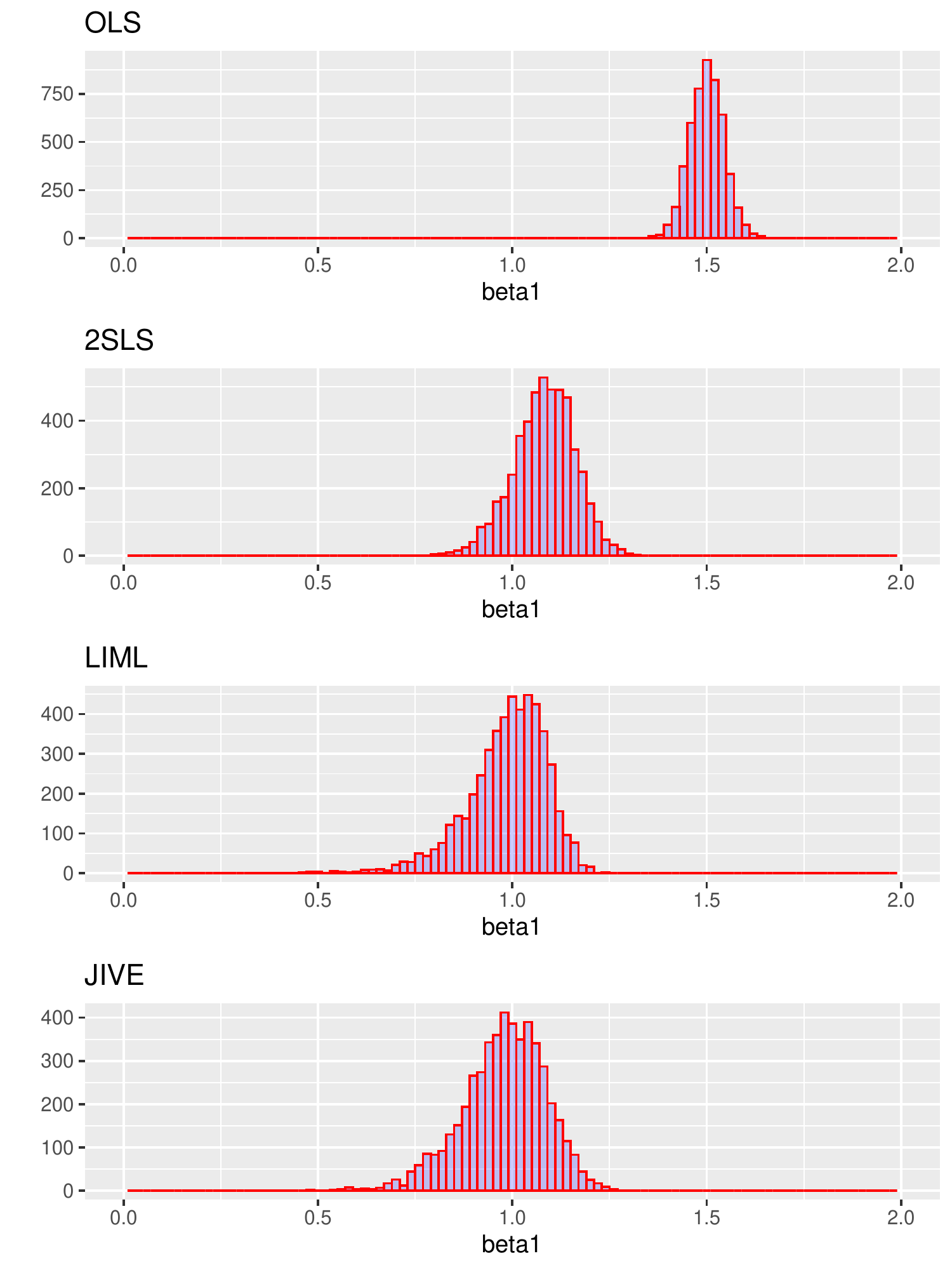}
\end{figure}

\begin{figure}[h]
\caption{\;Distribution of $\beta_1$ in Model 2.}
\centering
\includegraphics[width=0.9\textwidth]{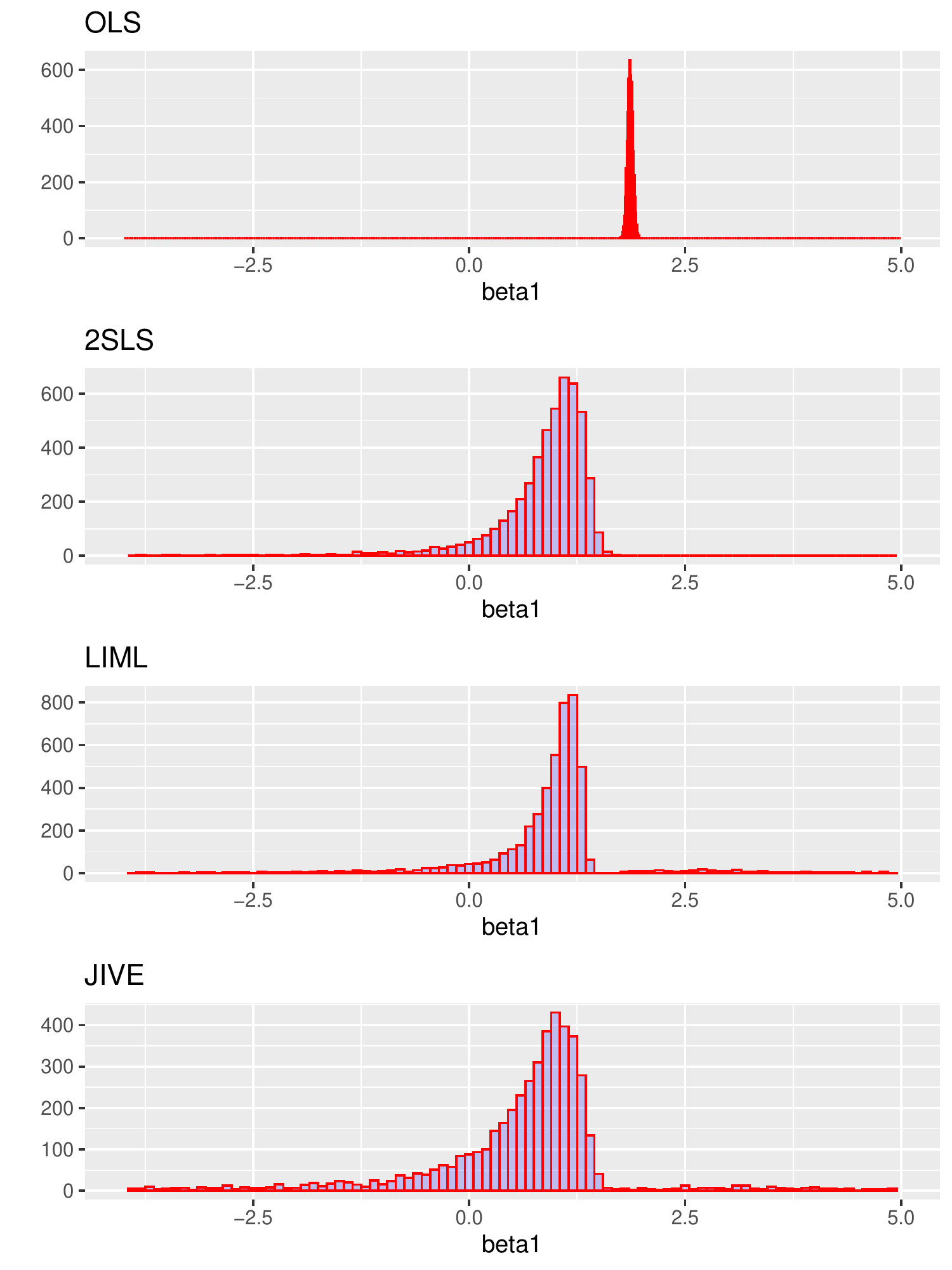}
\end{figure}

\begin{figure}[h]
\caption{\;Distribution of $\beta_1$ in Model 3.}
\centering
\includegraphics[width=0.9\textwidth]{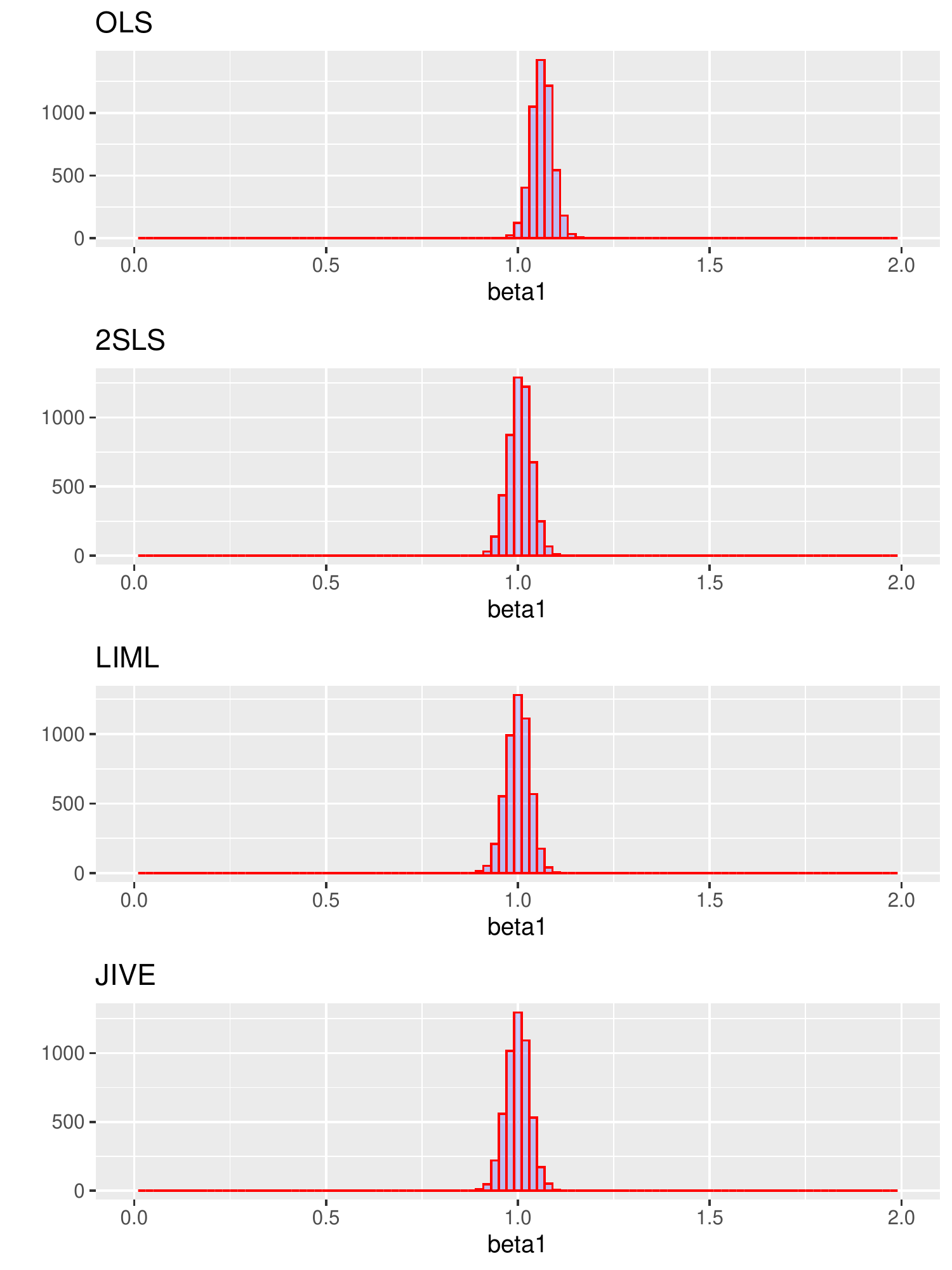}
\end{figure}

\begin{figure}[h]
\caption{\;Distribution of $\beta_1$ in Model 4.}
\centering
\includegraphics[width=0.9\textwidth]{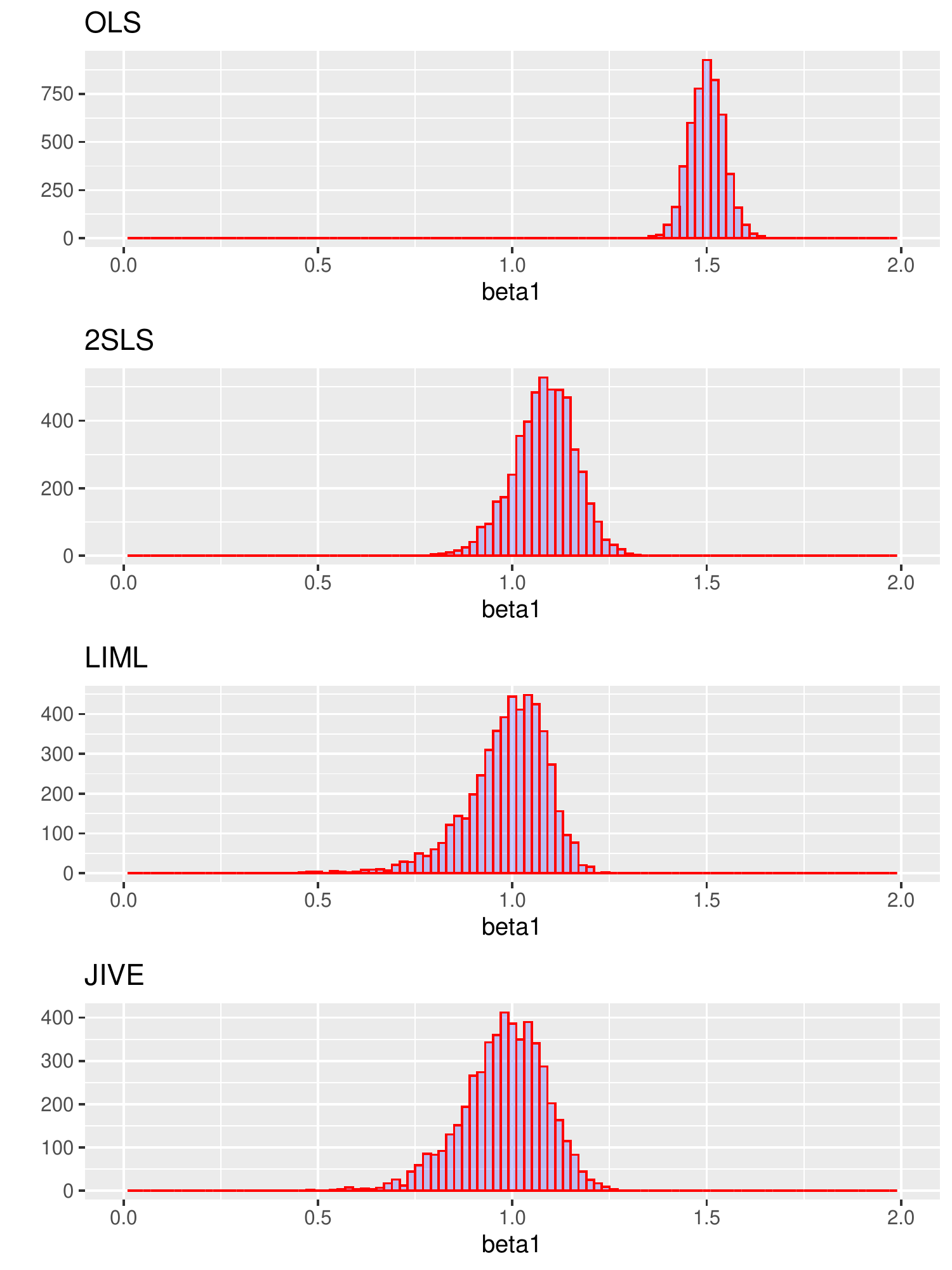}
\end{figure}

% Figure of real data application
%%%%%%%%%%%%%%%%%%%%%%%%%%%%%%%%%%%%%%%%%%%
\begin{figure}
\caption{Years of Completed Education and Quarter of Birth}
\label{fig:figure1}
\includegraphics[width=1.0\textwidth]{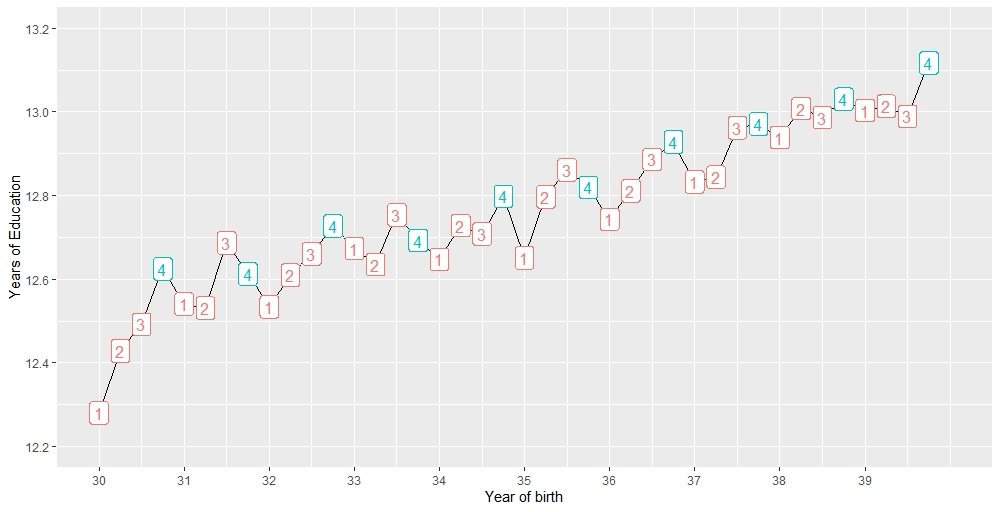}
\end{figure}

\begin{figure}
\caption{Quarter of Birth and Log Weekly Wages }
\label{fig:figure2}
\includegraphics[width=1.0\textwidth]{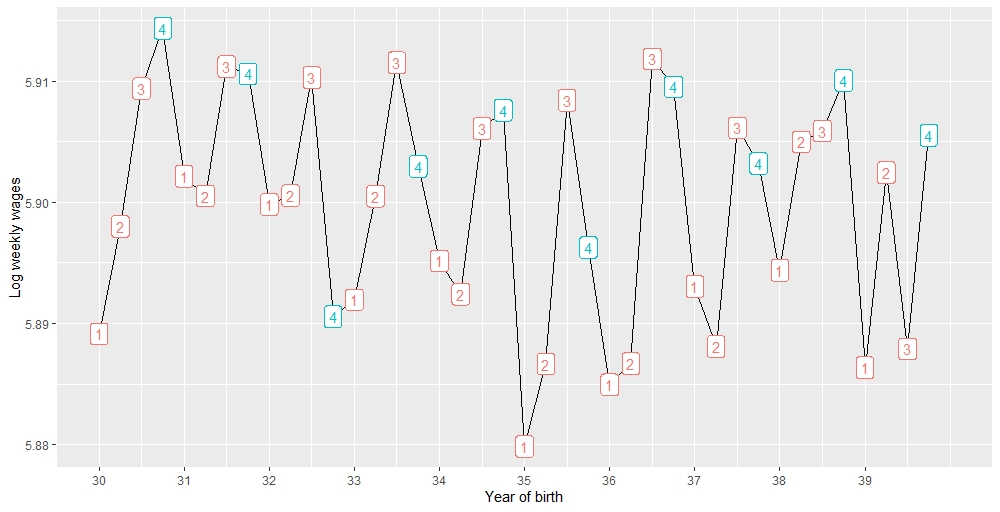}
\end{figure}

% Bibliography
%----------------------------------------------------------------------------------------
%\newpage % Includes a new page

\pagenumbering{roman} % Changes page numbering to roman page numbers
%\bibliography{literature}

\bibliography{main.bbl} % Add the filename of your bibliography
\bibliographystyle{unsrtnat}
% Defines your bibliography style

\end{document}